# Analysing Scientific Mobility and Collaboration in the Middle East and North Africa

Jamal El-Ouahi[1,2], Nicolas Robinson-García[3], Rodrigo Costas[1,4]

[1] *j.el.ouahi@cwts.leidenuniv.nl ; rcostas@cwts.leidenuniv.nl*
Centre for Science and Technology Studies (CWTS), Leiden University, Leiden, Netherlands

[2] Clarivate Analytics, Dubai Internet City, Dubai, United Arab Emirates

[3] *elrobinster@gmail.com*
Delft Institute of Applied Mathematics (DIAM), TU Delft, Netherlands

[4] DST-NRF Centre of Excellence in Scientometrics and Science, Technology and Innovation Policy, Stellenbosch University, Stellenbosch, South Africa

## Abstract

This study investigates the scientific mobility and international collaboration networks in the Middle East and North Africa (MENA) region between 2008 and 2017. By using affiliation metadata available in scientific publications, we analyse international scientific mobility flows and collaboration linkages. Three complementary approaches allow us to obtain a detailed characterization of scientific mobility. First, we uncover the main destinations and origins of mobile scholars for each country. Results reveal geographical, cultural and historical proximities. Cooperation programs also contribute to explain some of the observed flows. Second, we use the academic age. The average academic age of migrant scholars in MENA was about 12.4 years. The academic age group 6-to-10 years is the most common for both emigrant and immigrant scholars. Immigrants are relatively younger than emigrants, except for Iran, Palestine, Lebanon, and Turkey. Scholars who migrated to Gulf Cooperation Council countries, Jordan and Morocco were in average younger than emigrants by 1.5 year from the same countries. Third, we analyse gender differences. We observe a clear gender gap: Male scholars represent the largest group of migrants in MENA. We conclude discussing the policy relevance of the scientific mobility and collaboration aspects.

## Keywords

International mobility, collaboration, globalization, research policy, scientometrics indicators, Middle East and North Africa.

## 1. Introduction

In the words of the physicist Julius Robert Oppeinheimer, 'the best way to send information is to wrap it up in a person' (Oppenheimer, 1948). The mobility of highly proficient individuals is a key mechanism by which institutions acquire knowledge and stimulate creativity and innovation (Dokko & Rosenkopf, 2010; Mawdsley & Somaya, 2016; Palomeras & Melero, 2010; Singh & Agrawal, 2011; Slavova et al., 2015). They can serve as knowledge transmitters by transferring their prior knowledge to their receiving locations (Dokko et al., 2009; Jaffe et al., 1993). Additionally, they can intermediate connections with specialists known in prior locations (Breschi & Lissoni, 2009; Miguélez & Moreno, 2013; Singh, 2005). Scientists are no exception. Mobility has been described as a key aspect to improve scientific research (OECD, 2008; Scellato et al., 2015). Similarly, international collaboration promotes the production of high-quality knowledge (Wilsdon, 2011) and is indispensable to solve complex scientific problems (Sonnenwald, 2007). Scholars would usually produce higher-impact research when moving and collaborating internationally (Franceschet & Costantini, 2010; Gazni et al., 2012; Glanzel, 2001; Sugimoto et al., 2017; Van Raan, 1998).

Several authors have addressed the scientific mobility from a sociological and an economical perspective (Baldwin, 1970; Beine et al., 2008; K. E. Boulding, 1966; Kenneth E Boulding, 1966; Di Maria & Stryszowski, 2009; Hayek, 1945; Johnson, 1965; Kidd, 1965; Mountford, 1997). The most dominant concept of 'brain drain' appeared in the migration literature in the 1960s. First, it focused on the losses of highly skilled professionals from Europe, mainly the United Kingdom, to the United States described as the 'world's largest skills magnet' by Lowell (2003). It has been shown that the 'brain drain' had damaging effects for example in Eastern or Southern European countries (Ackers, 2005; Glytsos, 2010; Morano Foadi, 2006), or in Latin America (Times Higher Education, 2017) and Africa (OECD, 2015). Multiple innovative policy strategies even aimed at improving the 'brain drain' issue in regions such as in Asia (Krishna & Khadria, 1997; Song, 1997; Zweig, 1997). However, there is a clear uncertainty about the impact of international flows in academia. Other labels such as 'brain gain', 'brain circulation' for countries or 'brain transformation' for individuals are also commonly used. Cañibano and Woolley (2015) revised in detail the concept of 'brain drain' and its historical evolution. They also discussed the framework on 'diaspora knowledge network' introduced by (Meyer, 2001). Cañibano and Woolley (2015) concluded that these two frameworks, although useful, ignore structural and context dependent factors that affect mobility and its effects. Scott (2015) argues that labels currently used to discuss scientific mobility are out-of-date. He uses two broad frameworks to describe and analyse the mobility of academic staff. 'Hegemonic internationalization' is the dominant framework which focuses on migration flows from the 'periphery' to an evolving 'core'. The second framework labelled as 'fluid globalization' focuses on the emergence of global communities, social movements and issues of development. Scott (2015) concludes that the 'fluid globalization' framework may be more useful to understand the trends in scientific mobility. He describes the scientific mobility as a 'spectrum', from the deeply rooted to the highly mobile scientists, with most scholars standing in the middle of that spectrum. But his frameworks still focus on mobility flows from a 'periphery' to a single 'core', dominated by the West and increasingly evolving towards the East.

From a science policy perspective, collaboration and mobility studies improve the understanding of policy makers and research managers when assessing the scientific output of their countries or their organizations in a wider terrain of globalization. In the context of global mobility, nation states have developed their immigration policies to attract distinguished scholars and young researchers. On the one hand, collaboration and mobility are used as a means to integrate global scientific networks (Nerad, 2010). On the other hand, collaboration and mobility are means to internationalize national science systems. International mobility and collaboration are indeed perceived as two sides of internationalization, with the former being a trigger of the latter (Kato & Ando, 2017). While some countries depend on foreign-born scholars to preserve their scientific status (Levin & Stephan, 1999; Stephan & Levin, 2001), other countries consider mobility as a means to improve their national scientific capacities (Ackers, 2008), or to be considered as scientifically advanced countries (Kato & Ando, 2017). These cases are well positioned with the concept of internationalization perceived as the set of policies, programs and practices undertaken by academic systems, institutions and individuals 'to cope with globalization and to reap its benefits' (Altbach & Knight, 2007). Existing research provides evidence of positive effects of international mobility on the careers of scientists with the broadening of their networks (Netz et al., 2020).

It is only until recently that bibliometric methods have offered a plausible solution to macro-level analyses of international mobility (Laudel, 2003; Sugimoto et al., 2016). Computational advancements and especially the development of author name disambiguation algorithms, now allow tracking scientists mobility patterns based on changes in their affiliations in publications over time. The first macro studies on mobility using bibliometric methods were proposed by Henk Moed and colleagues (Moed et al., 2012; Moed & Halevi, 2014). These studies were mostly characterized by a brain drain/gain perspective, in which features such as multiple affiliation and cases of simultaneous affiliations were not specifically considered. To tackle this issue, Robinson-García et al. (2019) proposed a taxonomy of mobility types based on the persistence in time of scientists' linkage to countries. They distinguished between *migrants* and *travellers*. Migrants are characterized by having a cutting point in which they stop being affiliated to a country. Travelers maintain their linkage to a country, while adding other international affiliations (Ackers, 2005; Chinchilla-Rodríguez et al., 2018; Laudel, 2003; Robinson-García et al., 2018; Robinson-García et al., 2019; Sugimoto et al., 2016; Sugimoto et al., 2017). Among other advantages, bibliometric tracking of scientific mobility allows gaining access to mobility data in regions in which there is a lack of other sources of mobility information (e.g., surveys), as well as allowing diachronic analyses (Malakhov & Erkina, 2020; Miranda-González et al., 2020; Yurevich et al., 2020). Specific studies in different regions of the world and selected countries have been performed to better understand how they are integrated in the global network and how globalization affects specific geographical regions (Bernard et al., 2021; Subbotin & Aref, 2020; J. Wang et al., 2019; Y. Q. Wang et al., 2019; Zhao et al., 2021). Other studies have been conducted to develop individual-level migration data and key features of mobile researchers including patterns of migration by academic age, disciplines, and gender (Aref et al., 2019; Zhao et al., 2021). Such studies contribute to a better understanding of scientific mobility by policy makers and research managers in their countries or their institutions.

This paper contributes to Scott's frameworks on 'hegemonic internationalization' and 'fluid globalization' where we focus on regional mobility linkages to analyse the scientific mobility phenomenon in the Middle East and North Africa (MENA) region. MENA countries have made considerable investments in science and technology capacity to promote research and innovation (Schmoch et al., 2016; Shin et al., 2012; Siddiqi et al., 2016). Such investments specifically target at the internationalisation of their domestic research. For this, attraction of foreign talent is a key element. Some outcomes of such investment are already visible, with some of these countries experiencing a recent growth of scientific production (Cavacini, 2016; Gul et al., 2015; Hassan Al Marzouqi et al., 2019; Sarwar & Hassan, 2015). Several international experts' groups have regularly met to discuss the international migration and developments in some of the MENA countries (International Labour Office, 2009; League of Arab States, 2009; United Nations, 2002-2018). Few other reports and studies have also examined the migration of highly skilled workers in this specific region (Fargues, 2006; Özden, 2006; Unesco, 2015). The 'brain drain' framework is the main perspective in all these papers. Fargues (2006) and Özden (2006) also mentioned the poor quality or the lack of reliable migration data as well as the need of policies to enhance the benefits of migration for the development and the integration of the region. We also address the lack of reliable data. Özden (2006) presented the extent of the so-called 'brain drain' from MENA by using the dataset prepared by Docquier and Marfouk (2005). However, this data is limited to migration flows to OECD countries and ignores major destinations and origins for scholars in MENA.

In contrast to assuming MENA countries suffer from a brain drain in a more recent bibliometric study (Robinson-García et al., 2019), we observed that countries such as Qatar, Iraq, Saudi Arabia or the United Arab Emirates were world leaders in terms of relative attraction of foreign scientists. Clearly, a more nuanced theoretical perspective is needed to understand mobility in the MENA region.

In this paper, we focus on the MENA region aiming at better understanding the scientific mobility and collaboration in this region of the world. Specifically, we provide new ways to answer the questions that motivated earlier studies by pursuing the following research objectives:

i. To profile countries in the MENA region based on their mobile scientific workforce.
ii. To identify the main countries with which the MENA region interacts, distinguishing between origin and destinations of mobile scholars.
iii. To characterise the mobile scientific workforce in MENA countries based on their personal features. We focus specifically on their academic age (Nane, Larivière & Costas, 2017) and gender.
iv. To compare mobility and collaboration networks at the regional level.

The results of this study are expected to inform science policy makers in the MENA region, by providing them with additional evidence about the mobility patterns in the region, thus providing better and more contextualized interpretations to the policies regarding the mobility of the scholarly workforce in the MENA countries. Moreover, the results deployed in this study can also work as supporting evidence for policy makers from other countries and regions (e.g. Africa, EU, North America, Latin America, etc.) to understand the development of the MENA region regarding the internationalization of its workforce and its outcomes.

## 2. Data and Methods

### *2.1 Data collection*

In this study we use bibliometric data to track scientific mobility by identifying affiliation changes over time. We base our analyses on three Web of Science Core Collection indices (the Science Citation Index Expanded, the Social Sciences Citation Index and the Arts & Humanities Citation Index). We rely on an author name disambiguation algorithm to identify the complete publication history of scientists. Several algorithms have been proposed to perform such disambiguation (Backes, 2018; Caron & van Eck, 2014; Cota et al., 2007; D'Angelo & Van Eck, 2020; Mihaljević & Santamaría, 2021; Schulz et al., 2014; Torvik & Smalheiser, 2009). The most promising one is that by D'Angelo and Van Eck (2020), which filters and merges the results of the algorithm by Caron and Van Eck (2014) relying on an external source of information. This method achieves a precision of 96% and a recall of 96%. However, the existence of an external database is "crucial for the applicability of [their] approach" (D'Angelo & Van Eck, 2020, p. 904). This study presents a regional analysis of 22 countries, making it difficult to obtain such external source. Hence, we use the approach proposed by Caron and van Eck (2014). Although with lower precision (95%) and recall (90%), this algorithm is the unsupervised method producing the most promising results as shown by Tekles and Bornmann (2020).

We focus on the 2008-2017 period, as it is only possible to track affiliation changes in Web of Science since 2008, when authors and their affiliations started to be linked and recorded in the database. We identify a total of 22.6 million disambiguated authors who have published around 18.2 million distinct papers irrespective of the document types.

As per the World Bank (World Bank, October 2019), the MENA region is composed of 19 countries: Algeria, Bahrain, Djibouti, Egypt, Iran, Iraq, Jordan, Kuwait, Lebanon, Libya, Morocco, Oman, Palestine, Qatar, Saudi Arabia, Syria, Tunisia, United Arab Emirates and Yemen. We also included Afghanistan, Pakistan and Turkey as commonly included in the MENA region (also often called Middle East, North Africa, Afghanistan, and Pakistan (MENAP[1]) and Middle East, North Africa and Turkey (MENAT[2])). The dataset under study was comprised of 1,468,939 disambiguated authors who have contributed to 963,741 publications.

### *2.2 Indicators*

Table 1 lists the indicators we have used in our study as well as their definitions, how they are computed, and the types of data.

**Table 1. Indicators, definitions and calculations**

| *Indicator* | *Calculation* | *Type* |
|---|---|---|
| Academic Origin | Researcher's country affiliation on his first publication (Robinson-García et al., 2016; Sugimoto et al., 2017). | Demography |
| Academic Age | Age of the researcher's first publication (Nane et al., 2017). | |
| Gender | Gender of an author, inferred by an algorithm based on three different APIs: Genderize.io, Gender-guesser & Gender API which consider the first name of the author and the suspected country of origin. | |
| Mobility type | Taxonomy developed by Robinson-García et al. (2019) based on changes of author's affiliations. | Mobility |
| Average degree | The average degree measures the spread of influence across the network. Sum of all degrees divided by the number of nodes in a network (Hanneman & Riddle, 2005). | Network |
| Diameter | Maximum of distances between a pair of nodes in a network (De Nooy et al., 2018) | |
| Clustering coefficient | Proportion between the number of edges in the neighbourhood of a node and the number of potential edges in an entire weighted network (Barabási et al., 2002; Watts & Strogatz, 1998) | |
| Density | Degree of cohesion that exists among the vertices, determining whether a weighted network has a thin or thick consistency. Ratio of actual connections by number of potential connections (Wasserman & Faust, 1994) | |

[1] MENAP: https://www.imf.org/en/Publications/REO/MECA/Issues/2019/10/19/reo-menap-cca-1019#Sum
[2] MENAT : https://en.wikipedia.org/wiki/MENA

Although our study is limited to the 2008-2017 period, the academic age of a researcher is calculated based on his or her first publication which can be of course published before 2008.
In this study, we use the taxonomy developed by Robinson-García et al. (2019) which establishes the following mobility types:
1) *not mobile*, researchers who are always affiliated to the same country (e.g., country A);
2) *migrants*, those who leave at one point their country of first publication (e.g., they start in country A and are affiliated later with country B, and without further ties with country A). In this study we expand this typology by distinguishing at the country level between *emigrants* (for country A in our example before) and *immigrants* (for country B in our example before),
3) *travellers (directional)*, those who change countries but are linked to their country of origin throughout the study period (e.g., a researcher going from country A to A and B). We expand this typology to *outgoing* and *incoming travellers* (in the example before, A is the outgoing country, and B is the incoming country); and
4) *travellers (non-directional)*, researchers who are always linked to the same set of countries and hence we cannot establish the direction of movement (e.g., researchers affiliated to countries A and B in all the publications).

As a result of the above, we apply five final typologies of mobility to characterize the workforce of each country: not mobile, emigrant, immigrant, outgoing travellers and incoming travellers.

**Table 2. Researchers by mobility type in MENA (2008-2017)**

| *Mobility type* | *Total Share* | *Mobility Share* | *Total* |
|---|---|---|---|
| ***Not Mobile*** | 84.7% | | 1,244,858 |
| ***Mobile*** | 12.1% | 100% | 177,027 |
| *Migrants* | 3.3% | 27% | 48,134 |
| *Traveller (directional)* | 5.7% | 47% | 83,323 |
| *Traveller(non-directional)* | 3.1% | 26% | 45,570 |
| ***Insufficient information*** | 3.2% | | 47,054 |
| All | 100% | - | **1,468,939** |

Table 2 shows the number of researchers for each mobility type during the 2008-2017 period for the whole MENA region. Most researchers (84.7%) have not shown any sign of international mobility whereas around 12% have. Mobile scholars are mainly *Travellers (directional)*, representing 5.7% of the researchers under study. *Migrant* is the second most common type of mobility in MENA (3.3%), followed closely by *Traveller (non-directional)* (3.1%).

As noted by Robinson-García et al. (2019), the share of researchers by mobility type increases as the number of publications by researcher increases. However, the same authors observed an exception for non-directional travelers: more than half of the researchers assigned to this typology have published one or two papers. This led the authors to consider that this group may be affected by the potential errors derived from the disambiguation algorithm used in our study, which tends to split authors when the probability of publications belonging to the same author is low. To prevent from such limitation, in this study we exclude non-directional travelersfrom our analyses. It is worth noting 47,054 authors do not hold enough information either from the author disambiguation algorithm or from the mobility taxonomy which requires the publication of at least two publications to track the change of affiliations. These scientists were also excluded from our analyses.

Figure 1 shows the number of mobile researchers per country along with their mobility type. Considering the relatively low numbers of mobile authors in Djibouti, Bahrain, Afghanistan, Palestine and Yemen, these 5 countries were excluded from our dataset.

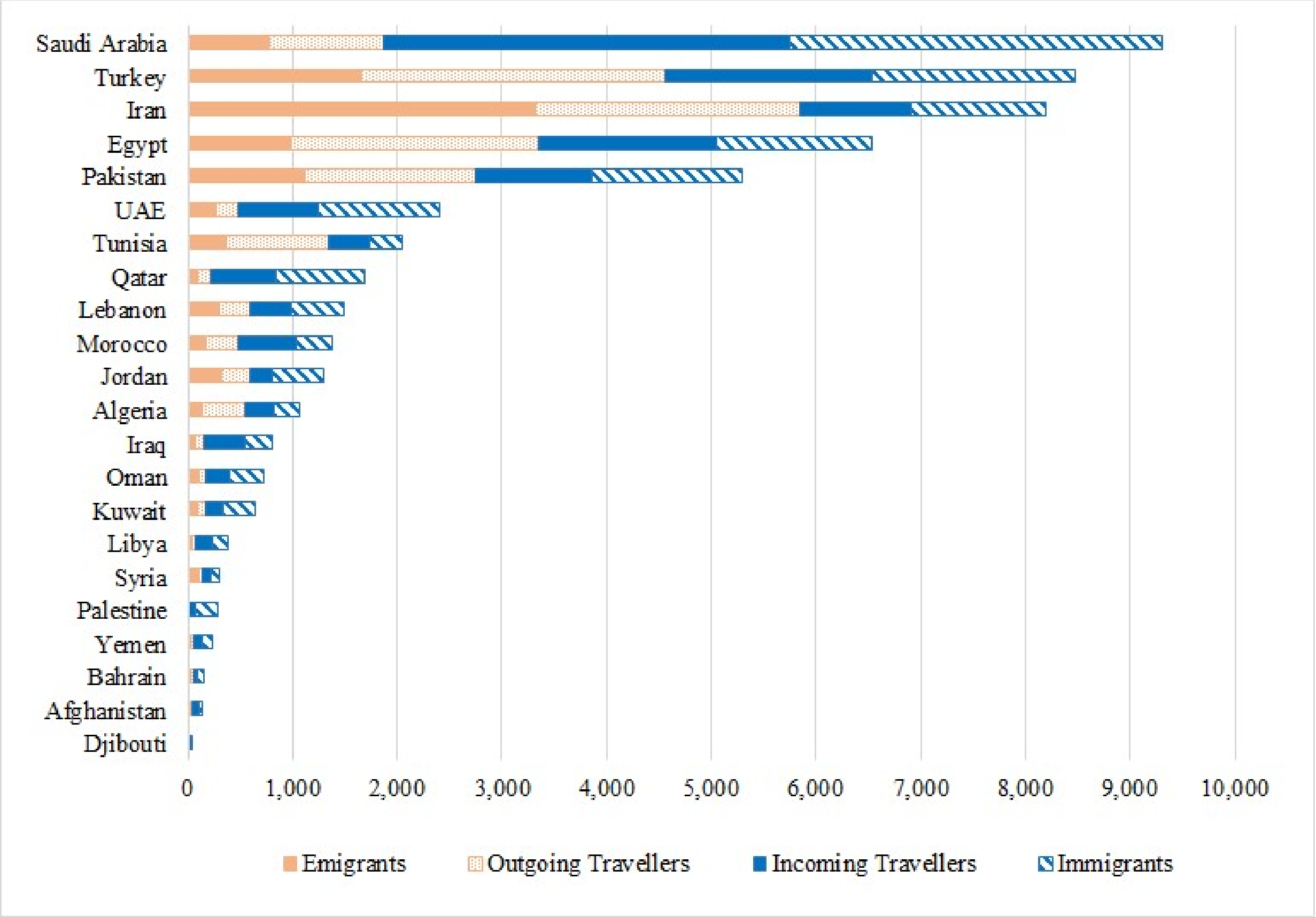

**Figure 1. Number of mobile directional scholars per country and mobility type (2008 - 2017).**

To infer a gender to authors, we follow the same strategy as the one employed in the 2019 edition of the Leiden Ranking[3].We infer a gender based on researchers' first name and their suspected country of origin. If no gender can be inferred, it is then considered unknown. The process is the following. First, for each author, one or more countries of origin are determined. In a publication, each author is linked to an affiliation which includes an address with a country. If the country of the author in his or her first publication is the same as the country the author is most often associated with in his or her set of papers, we then consider this country as the author's country of origin. Otherwise, we consider there is not enough evidence to define a single country of origin. All countries to which an author is linked are considered to be countries of origin. Then we used three tools to infer a gender: Gender API (gender-api.com), Gender Guesser (pypi.org/project/gender-guesser), and Genderize.io (genderize.io). It has been shown Gender API performs better as evaluated in a previous study (Santamaría & Mihaljević, 2018). The first name of the author combined with a country of origin were provided as inputs to these tools. This approach was applied to 24.6 million authors in the Web of Science with a confidence level of 90%. For 44% of them, a male gender was inferred. A female gender was inferred for 25% of the authors. For the remaining 32% of the authors, no gender could be inferred and have been labelled as N/A when the gender is unknown. We should keep in mind that these shares vary from country to country as shown in Appendix A. A male gender was inferred to 57% of the disambiguated authors affiliated to a MENA country during the study period. For 33% of them, a female gender was inferred while no gender could be inferred to the remaining 10%.

[3] https://www.leidenranking.com/information/indicators#gender-indicators

*2.3 Network analysis*

We constructed co-authorship networks as a proxy to examine collaboration patterns within the scientific community in MENA. These networks are presented at the national level, with countries represented by nodes and the number of co-authored papers by vertices. Two countries are connected by an edge when at least one scholar from country A has co-authored a paper with a scientist from country B.

In the case of the mobility networks, the methodology varies slightly. Here, edges represent the number of researchers who have been affiliated at any given point in time within the study period between countries A and B. Two countries are connected by an edge when at least one scholar has a mobility event from a country to another. Network visualisations were created using VOSviewer (van Eck & Waltman, 2009).

*2.3 Limitations of bibliometric approaches for mobility*

It is important to acknowledge upfront that there are several limitations to the methods we used. First, our methods rely mainly on tracking the changes in authors affiliations to measure the mobility. Thus, researchers with low number of papers would most probably be underrepresented (Abramo et al., 2011). Second, certain types of mobility events, such as short-term stays, are not necessarily translated into publications. A third limitation is due to the coverage in Web of Science, thus limiting our study to publications in indexed journals. Fourthly, the author-name disambiguation algorithm we used (Caron & van Eck, 2014) uses rule-based scoring and clustering based on bibliographic information such as author name, e-mail address, affiliation, publication source and citation information. The method used is conservative as it values precision over recall. If there is not enough evidence to group publications together, they will be grouped in separate clusters. Errors in publications coupling might occur for several reasons. For example, an author with high frequency of affiliation change might be clustered into several different 'authors' by the algorithm. To a lesser extent, this problem might also apply to authors who did not change their affiliations. For many authors, the algorithm splits up the publications under multiple author identities. Typically, there is one dominant identity that covers most of the papers and few separate identities that include only one or two publications. These are considered as artefacts of the disambiguation of the algorithm and are excluded from our study. Nevertheless, with these limitations in mind, Sugimoto et al. (2017) discussed to some extent the validity of the approach used to identify international mobility by comparing the mobility algorithms with affiliation data recorded in Open Researchers and Contributor ID (ORCID) public data file, finding that about 63% of researchers mobile in ORCID were also identified by bibliometric means, supporting the relevance of bibliometrics but also highlighting the relative conservative perspective of the bibliometric approach. To assess the accuracy of the approach for disambiguating authors, we compared our dataset with the 2020 ORCID public data file that we consider as a reference dataset.

In this file, the registered researchers are uniquely associated with their scientific oeuvre. We used this information to verify the accuracy of the disambiguation algorithm. We first matched the publications of our dataset with those available in ORCID by using unique identifiers such as the DOI, the Web of Science Accession Number or the PubMed ID. For the matched records, we examined the authors disambiguated by the algorithm with the authors in ORCID by analysing the last name and the first forename initial. If the name strings from both sources match, we assume they refer to the same author. We also use the e-mail address as an additional information to match the author names. 6,459 disambiguated authors were associated with an ORCID. Table 3 reports the number and the percentage of correct and incorrect (i.e., researchers with more than one disambiguated cluster) disambiguation. 91.1% of the disambiguated authors were correctly matched with one ORCID, while 8.9% authors in the ORCID public file were split into multiple disambiguated authors by the algorithm.

**Table 3. Statistics for our approach to author disambiguation versus ORCID records**

| | **Disambiguation Algorithm** | |
|---|---|---|
| | **Correct** | **Incorrect** |
| **ORCID** | 5,884 (91.1%) | 575 (8.9%) |

Finally, the algorithm we used to infer the gender of authors is of course not perfect and we should keep these limitations in mind when analysing the results. Overall, the limitations discussed above point that we are most likely underestimating the true mobility that we are measuring, and therefore we are taking a quite conservative approach, in which we expect a high precision in what is captured (i.e., the mobility events are correct in the framework of this paper), but not all mobility events can always be properly identified.

## 3. Results

In this section, we present the main findings of the study. First, we offer an overview on the number of identified scientists by country as well as the proportion they represent by mobility types at the regional level. Next, the mobility profiles of each country in MENA are presented, followed by an analysis of the mobility flows. Then, we focus on the gender and the academic age of mobile scholars. Finally, we compare the mobility and the collaboration networks.

### *3.1 General results and countries profiles*

In Figure 2, we summarize the number of disambiguated researchers per country as well as the papers published during the study period. Authors affiliated to Iranian institutions show the highest rate of publications per researcher, followed by scholars in Turkey and Tunisia.

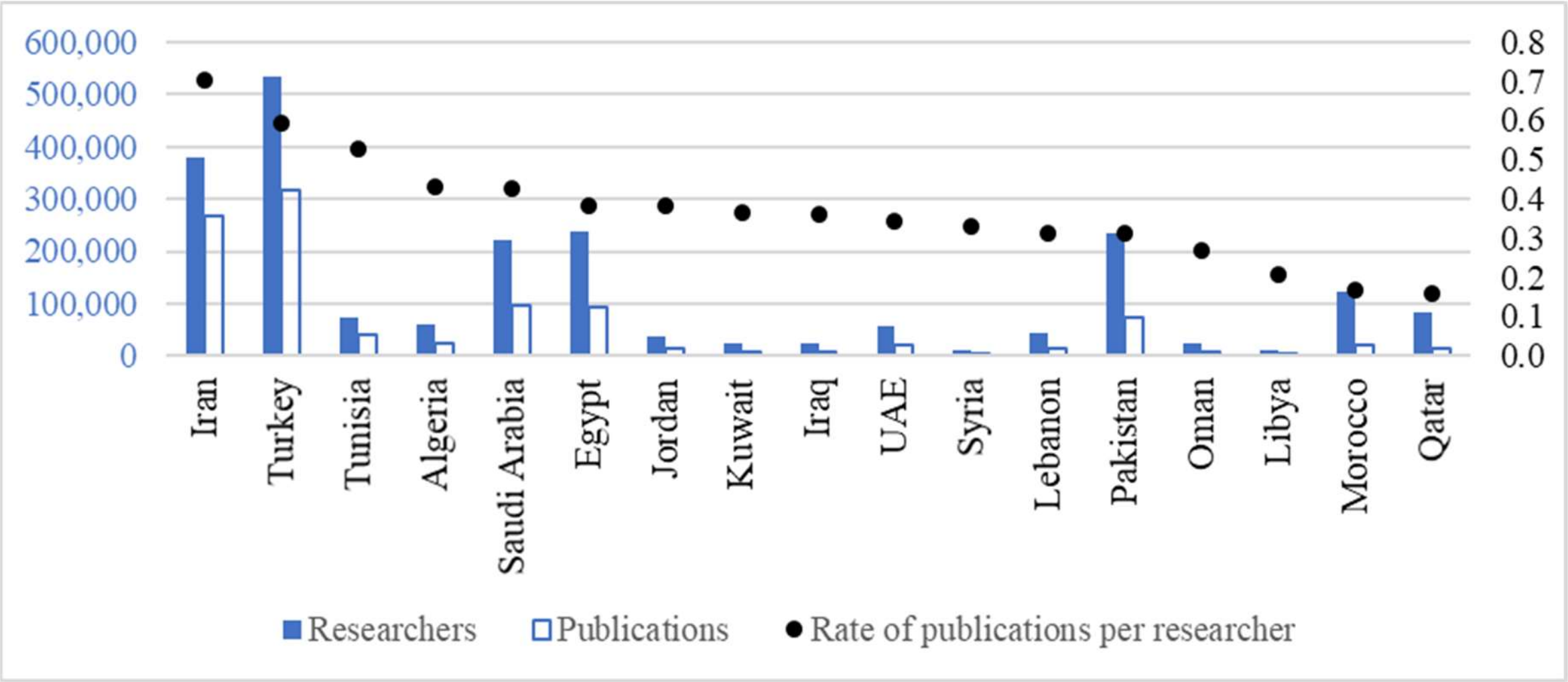

**Figure 2. Number of researchers, publications and rate of publications per researcher by country (2008-2017)**

We develop country profiles of the MENA region based on the mobile scientific workforce identified. In Figure 3, we report the in-migration and the out-migration per country. Saudi Arabia, United Arab Emirates, Qatar, Kuwait and Oman, part of the Gulf Cooperation Council (GCC), all have higher rates of incoming scholars (~79%) than outgoing. These five countries are the only MENA countries having a High-Income level as per the World Bank (June 2019). To a lesser degree, Morocco, Lebanon, Syria and Jordan also have a higher share of incoming scientists (~63%) than outgoing ones.

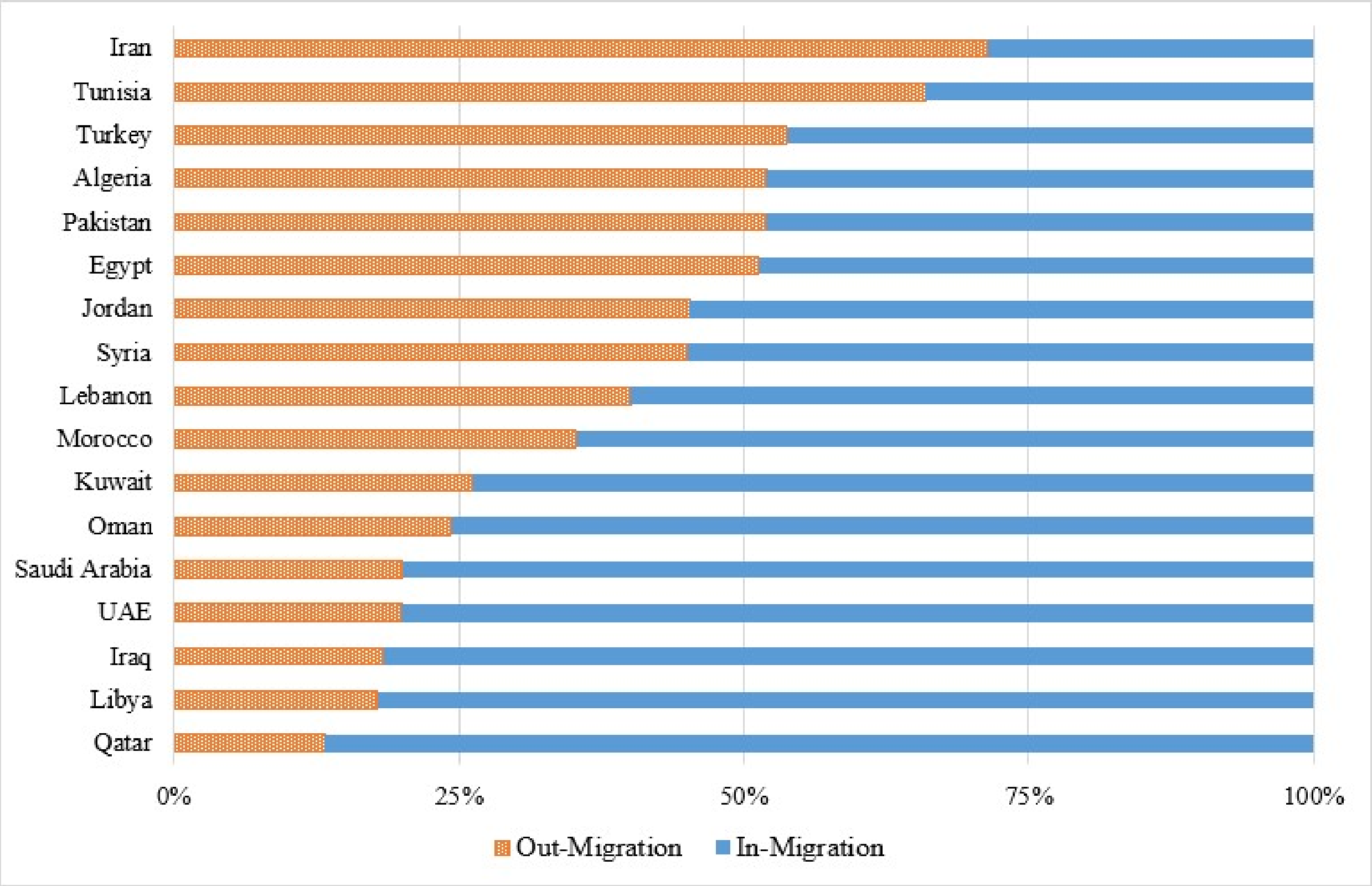

**Figure 3. Share of mobile directional researchers by mobility type per country (2008 - 2017).**

Several countries have larger shares of outgoing scholars (either as emigrants or outgoing travellers) than incoming. Iran and Tunisia have the highest shares of outgoing scholars respectively 71% and 66%. Iran and Syria show the highest rate of emigrant scientists. Turkey, Egypt, Algeria and Pakistan have similar shares, where around 52% of their mobile researchers are emigrants or outgoing travellers. Qatar, Saudi Arabia and United Arab Emirates are getting the most influx of researchers compared to very small outflows. On the other hand, Syria, Jordan, Iran and Lebanon have the highest rate of outgoing flows. When comparing the shares of emigrants and immigrants, Iran, Tunisia and Syria are the only countries which show an overall deficit of researchers.

*3.2 Mobility networks at the regional and country levels*

Next, we look at the flows of scholars moving from and to MENA countries. Figure 4 offers an overview of the mobility phenomenon for MENA scholars. All origins and destinations of scientists affiliated to a MENA country at some point in time between 2008 and 2017 are grouped by continent. It is worth noting that the MENA region is composed by countries located in North Africa and West Asia.

Figure 4 shows the flows of mobile scientists at the regional level. Each node or vertical bar represents a region. The size of the flow between two nodes represents the number of scientists who have moved from a region to another. This figure shows that the MENA region has overall more inbound than outbound flows. For all other regions, the inbound and outbound flows have relatively the same size.

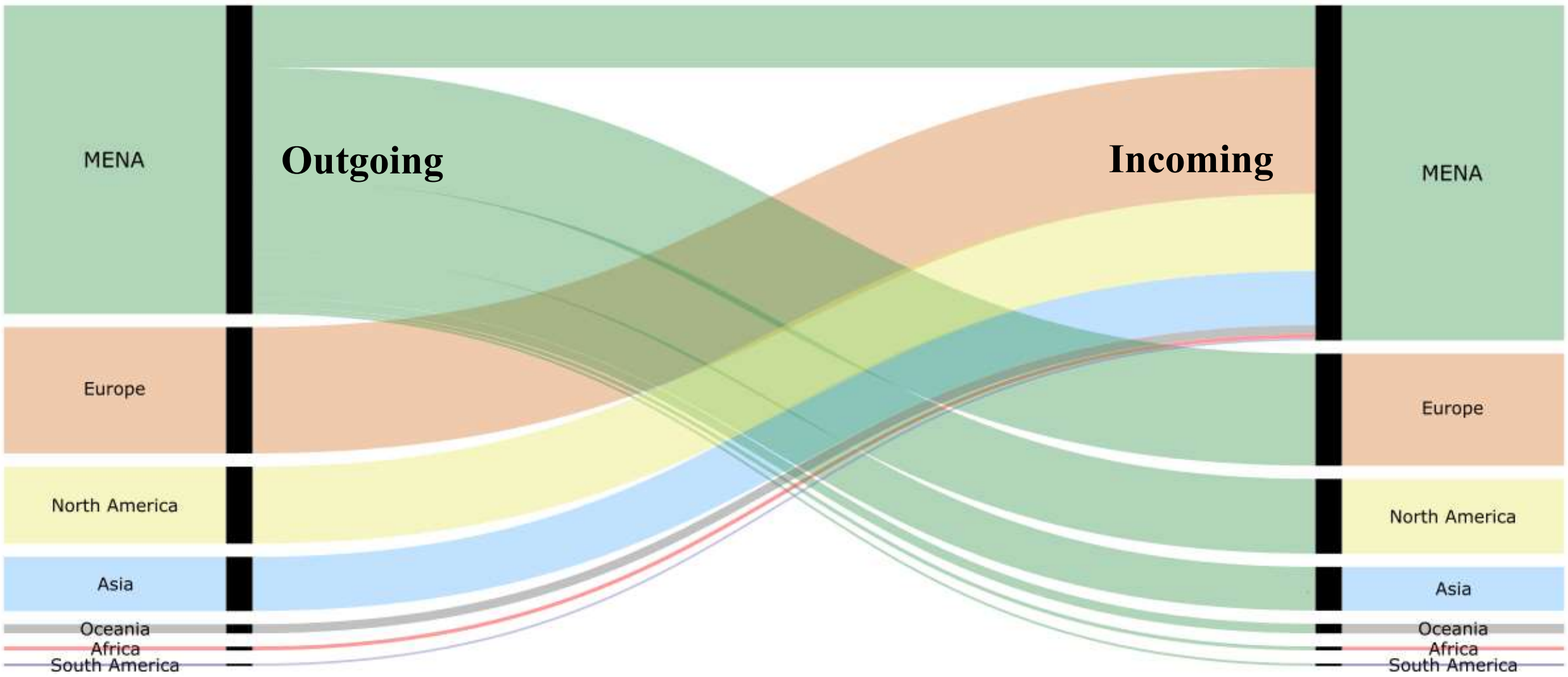

**Figure 4. MENA Mobility flows at the regional level (2008 - 2017).**

The MENA region is highly connected with Europe based on the number of mobile scientists. Europe is indeed the first mobility destination and origin with 37% of the flows from/to MENA, followed by North America (24%), MENA (20%) and Asia (16%). These findings suggest a relatively high level of intra-MENA flows. Oceania, Africa, and South America show a much lower circulation of scholars (less than 3%).

Next, we analyse inter-countries flows. Figure 5 shows the mobility flows of scholars moving from and to the MENA region by countries labelled with their ISO Alpha-3 Codes. Only countries with more than 350 mobile scientists between 2008 and 2017 are shown. United States, France, United Kingdom, Germany, Canada, China, Malaysia, Italy, Japan, and Australia are the main non-MENA destinations and origins. Furthermore, figure 4 shows that flows are not only limited to scholars moving from developing countries to developed countries. When analysing the origins and destinations of mobile scholars, United States appears to be the most common destination and origin for migrant scholars who were affiliated to an institution in the MENA region between 2008 and 2017.

When looking at specific MENA countries, some cases stand out. For example, France is the preferred destination for scholars originating from its former colonies in MENA, specifically Morocco, Algeria and Tunisia. North African countries have also strong ties with other countries in Europe such as Spain, Germany, Switzerland and Netherlands. United Kingdom is one of the preferred destinations for GCC countries such as Saudi Arabia, the United Arab Emirates and Qatar. Scholars from Egypt and Jordan have mostly migrated to Saudi Arabia, ahead of United States. Researchers from Pakistan migrate mainly from and to China. Iraq and, to a lesser extent, Iran have major flows from and to Malaysia. In the case of Iran, it is worth reminding that the political sanctions from the United States have had an impact on the scientific international collaboration (Kokabisaghi et al., 2019). For example, Iranian scholars have been denied opportunities to attend international scientific meetings during periods of sanctions. The blockade of the Iranian Rial exchange has prevented Iranian researchers to pay publication, conference registration and membership fees in foreign currency.

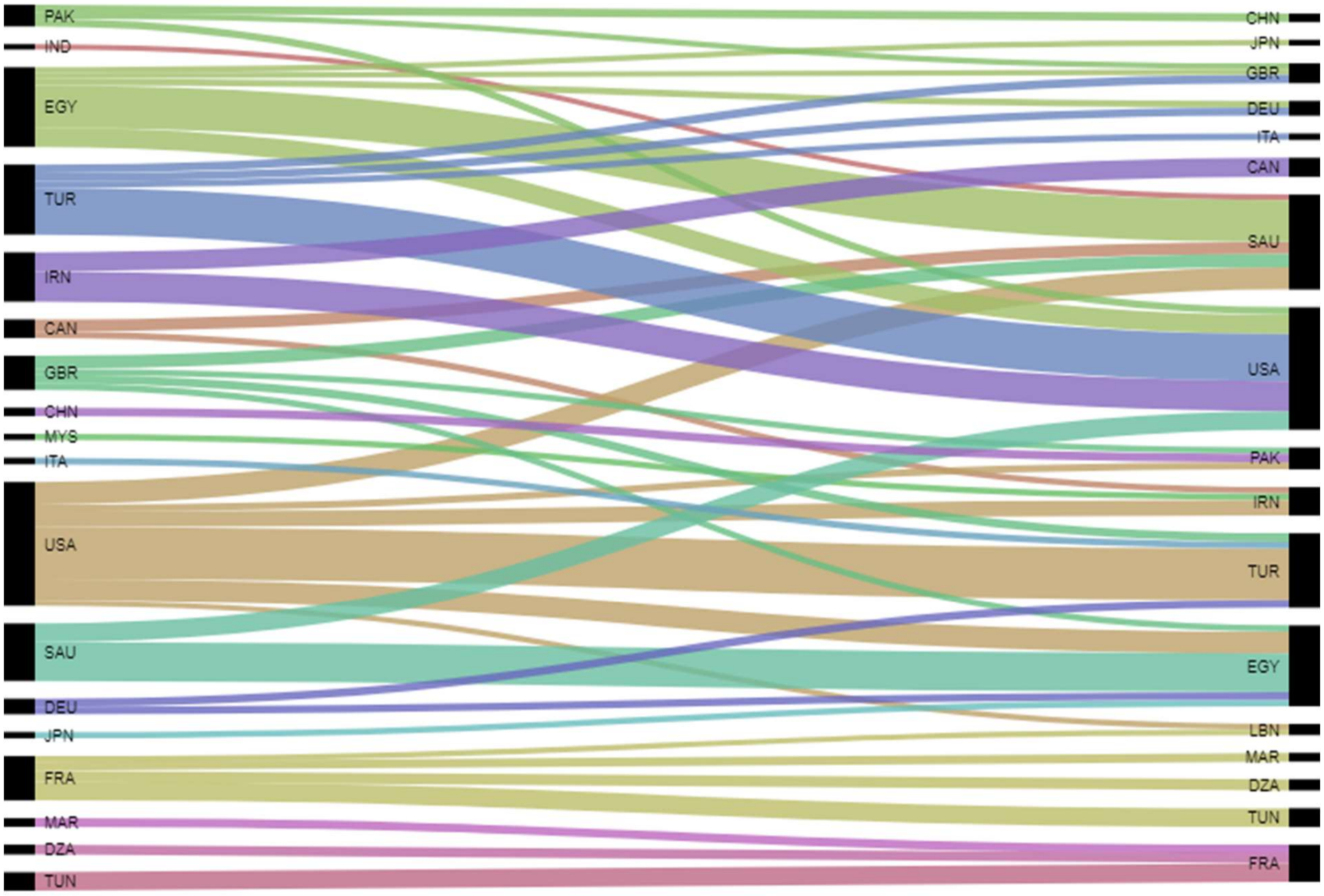

**Figure 5. Mobility flows for scholars from/to MENA countries (2008 - 2017).**

We see within the top 15 destinations/origins of MENA migrant scholars that, except for Pakistan and Iran, we can already find some countries outside of the region. Some of these cases could be explained by geographical, cultural, historical, linguistic and socio-political proximities (Scott, 2015).

### *3.3 Individual characteristics of the migrant scientific workforce: Gender and Academic Age*

We now investigate the personal features of the migrant scholars by analysing their distribution by academic age and gender. In terms of mobility, the migrant scholars represent the most policy-relevant group as they change their countries of affiliation whereas the travellers keep an affiliation with their suspected countries of origin. Figure 6 shows a population pyramid based on the average age of Emigrant and Immigrant scholars in the MENA region.

The average academic age of migrant scholars in MENA between 2008 and 2017 was 12.39 years. For the whole MENA region, Immigrants have an average academic age of 12.5 years versus 12.3 for the Emigrants. For most countries, the immigrants are relatively younger than emigrants, except for Iran, Palestine, Lebanon and Turkey. The academic age group '6 – 10' years is the most common for both emigrant and immigrant scholars. This group represent around 42% of all the migrants. '11 – 15' is the second age group, representing 32% of the migrant scientists. Migrant scholars with an academic age between 16 and 20 years correspond to 10% of migrants. Other age groups represented less than 6%. Scholars who migrated to GCC countries, Jordan and Morocco were in average younger than emigrants by 1.5 year from the same countries as represented in Appendix B. In this appendix, we focus only on emigrants and immigrants for countries where more than 1,000 mobile researchers have been identified.

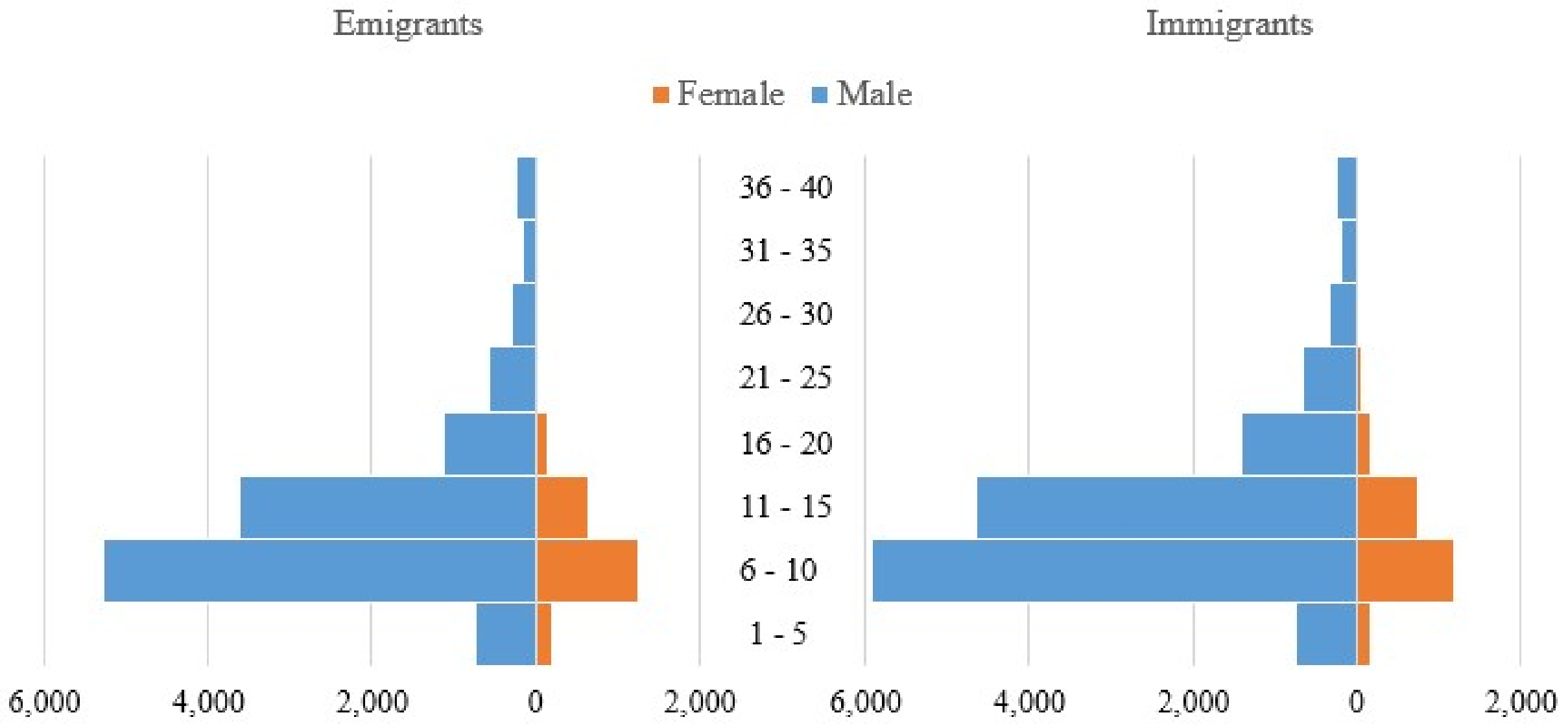

**Figure 6. Population pyramid of migrant scholars in MENA (2008 - 2017).**

Male scholars represent 66% of all migrants in MENA and female authors account for 12%. For the remaining 22%, gender was not reliably identified. These shares are similar when comparing between emigrants and immigrants. However, we observe differences by country (see Appendix B.) Tunisia and Lebanon have the highest shares of female emigrants, 22% and 21% respectively. They are followed by Turkey, Algeria, Morocco and Iran with around 17% of female scholars. Pakistan and Egypt have a share of around 11% of female migrant scientists. In the remaining countries, female authors represent shares below 10% with the lowest shares (about 7%) reached in Iraq, Saudi Arabia, Syria and Libya.

Figure 7 shows on the X-axis that almost all countries in MENA are dominated by male researchers. The only countries for which the gender ratio is close to 1 are Tunisia, Lebanon and Turkey. The average male-to-female ratio for Iraq, Saudi Arabia, Jordan, UAE, Qatar and Pakistan exceeds 3. In the same figure, we also examine the gender ratios among the migrant researchers for each country, and then compare them to the corresponding ratios among all researchers.

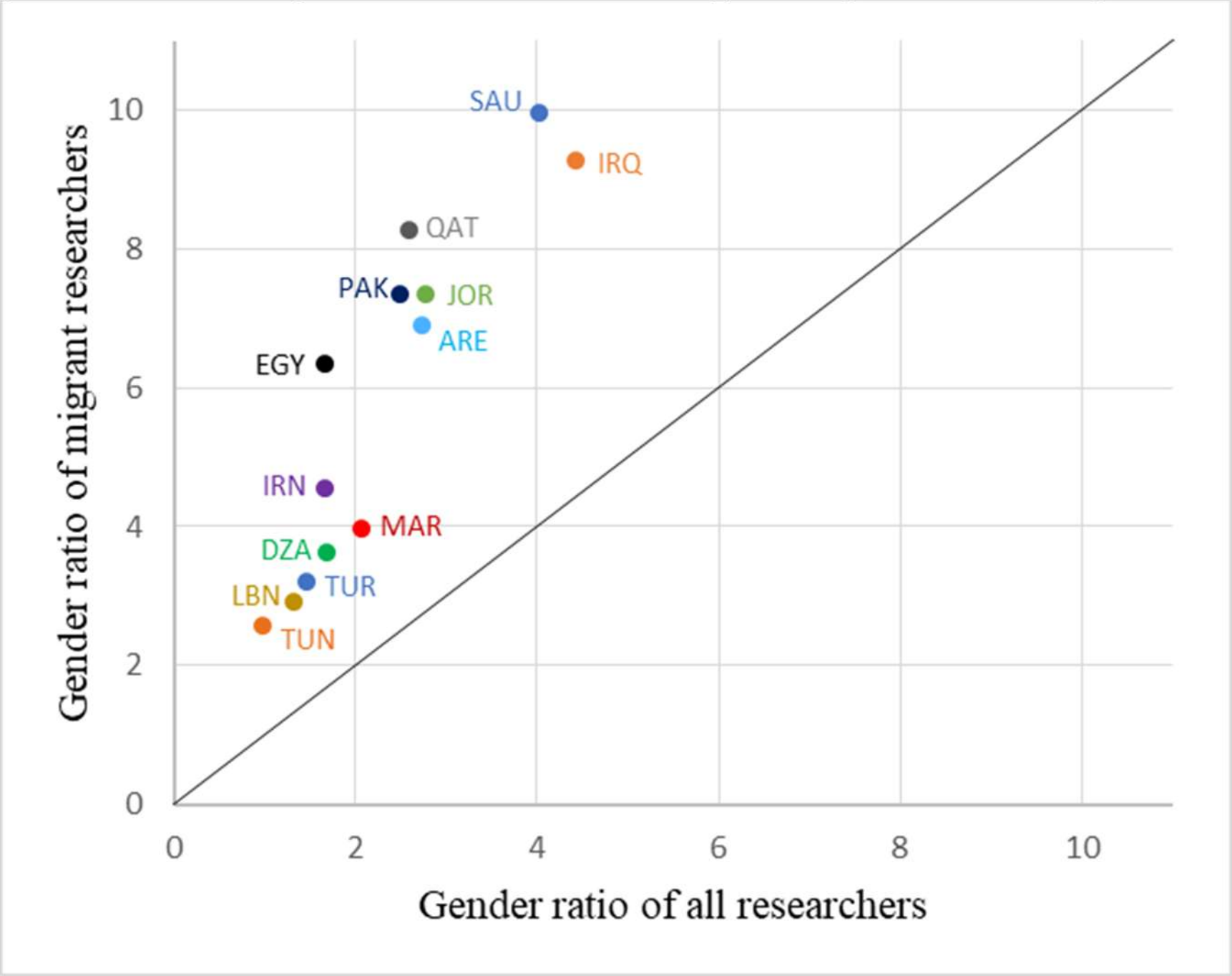

**Figure 7. Male-to-female gender ratios of migrants and all researchers by country**

We notice a clear gender gap in terms of scientific mobility. In all countries, the gender disparity is more severe among the migrant researchers. The male-to-female ratio among migrant researchers is on average 2.5 times higher than the male-to-female ratio for all researchers.

*3.4 Comparison of collaboration and mobility networks*

Following, we compare the international scientific collaboration and mobility networks of MENA countries. Figure 8 shows the MENA international collaboration network. Saudi Arabia, Iran, Egypt and Turkey drive most of the international cooperation within the region. However, the partnerships of these three countries seem to vary. While Saudi Arabia, Iran and Egypt show stronger collaboration links with some Asian countries, Turkey shows strong collaboration linkages with several European countries such as Germany and France. Our findings are also consistent with the results previously published in the Towards 2030 report (Unesco, 2015): Iran has strong collaboration ties with developing countries. Malaysia is among the top 10 collaborators, but Iran has a low share of papers with a foreign co-author. Still, we must note the role of United States and United Kingdom as important actors within the network driving strong collaboration linkages with most of the MENA countries.

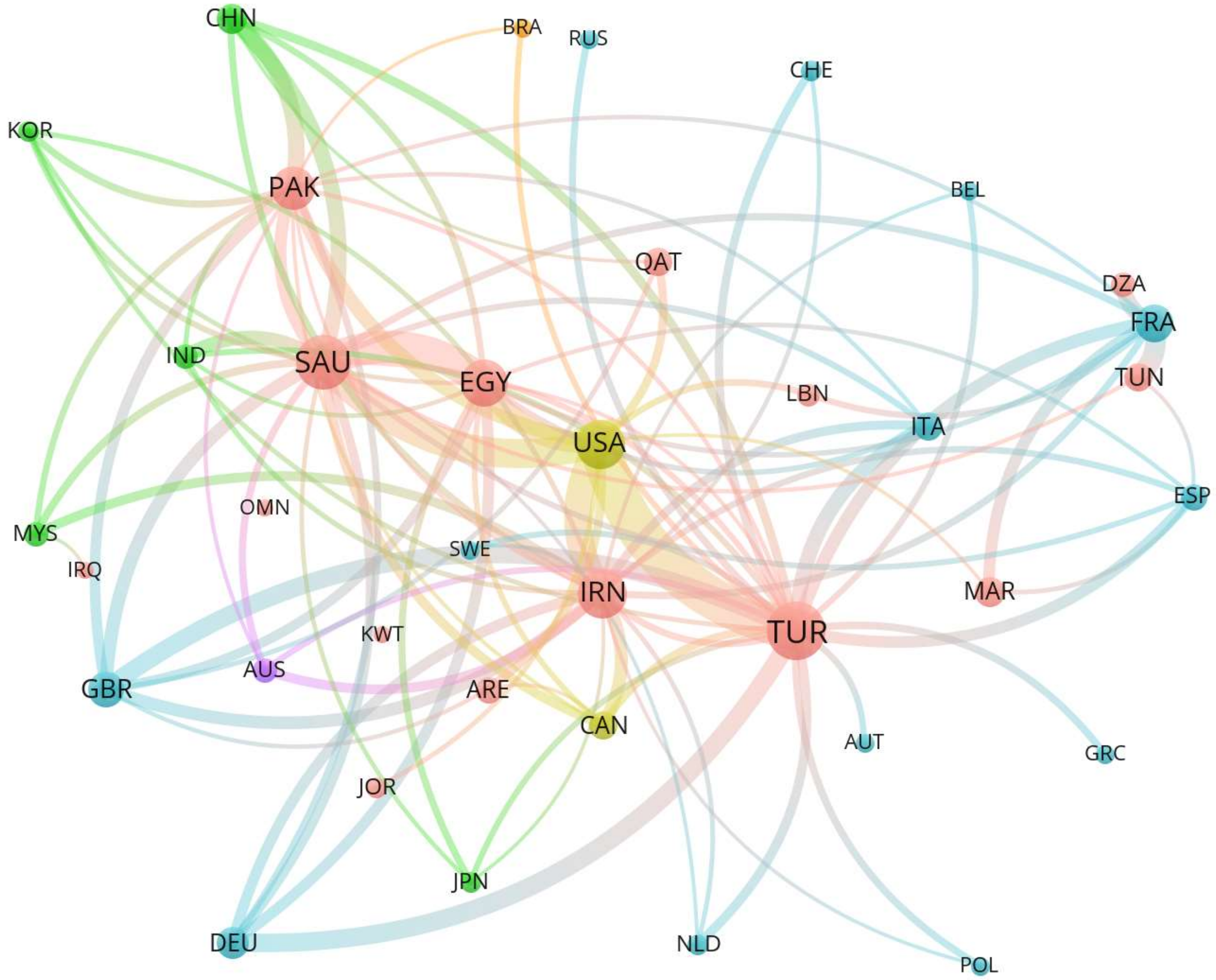

**Figure 8. Main countries and links in the MENA Collaboration Network (2008-2017). Co-authorship relations with at least one author from a MENA country and at least 100 co-publications at the country level are included. For readability reasons, we show here the 100 strongest links between the countries. Colours of nodes represent world regions**

Few classical measures of network analysis are listed in Table 4 to describe the structures of the collaboration and mobility networks. The number of nodes or countries linked to other countries are different in the networks. Scholars in MENA migrate or travel to less countries they collaborate with. The number of edges represent the number of links between countries. The lower number of links in the mobility network is also reflected in the lower density. The density helps us to evaluate the degree of cohesion that exists between countries. International co-authorships we used to measure scientific collaborations tend to be more frequent than international mobility. The mobility network has a thinner consistency than the collaboration network in terms on affinity between countries (Wasserman & Faust, 1994).

**Table 4. Structural measures of the MENA mobility and collaboration networks (2008-2017)**

| **Structural Measure** | **Mobility** | **Collaboration** |
|---|---|---|
| *Number of vertices* | *176* | *215* |
| *Number of edges* | *1,335* | *3,124* |
| *Density* | *0.09* | *0.14* |
| *Average degree* | *2.07* | *1.87* |
| *Diameter* | *4.00* | *3.00* |
| *Clustering coefficient* | *0.29* | *0.24* |
| *Assortativity* | *-0.72* | *-0.88* |

MENA countries tend to collaborate at the international level to a higher degree than they exchange human resources. Indeed, international co-authorships we used to measure scientific collaborations tend to be more frequent than international mobility which is a rarer event. The assortativity coefficient is the Pearson correlation coefficient of degree between pairs of linked nodes (Newman, 2002). Positive values indicate a correlation between nodes of similar degree, while negative values indicate relationships between nodes of different degree. The negative values of assortativity for both networks indicate that MENA countries with small degrees tend to connect with countries with higher degrees.

Next, we measured the degree and closeness for each MENA country in mobility and collaboration networks during the study period. Table 5 lists centrality measures for each node to describe the role of each country in collaboration and mobility.

**Table 5. Centrality measures of MENA countries in collaboration and mobility (2008-2017)**

| | *Mobility* | | *Collaboration* | |
|---|---|---|---|---|
| *Country* | *Degree* | *Closeness* | *Degree* | *Closeness* |
| Algeria | 64 | 0.61 | 169 | 0.83 |
| Bahrain | 40 | 0.56 | 150 | 0.77 |
| Egypt | 101 | 0.70 | 192 | 0.91 |
| Iran | 95 | 0.68 | 193 | 0.91 |
| Iraq | 61 | 0.59 | 154 | 0.78 |
| Jordan | 59 | 0.60 | 174 | 0.84 |
| Kuwait | 68 | 0.61 | 163 | 0.81 |
| Lebanon | 73 | 0.63 | 175 | 0.85 |
| Libya | 45 | 0.55 | 151 | 0.77 |
| Morocco | 85 | 0.66 | 189 | 0.90 |
| Oman | 86 | 0.66 | 169 | 0.83 |
| Pakistan | 104 | 0.71 | 187 | 0.89 |
| Palestine | 29 | 0.51 | 152 | 0.78 |
| Qatar | 89 | 0.67 | 171 | 0.83 |
| Saudi Arabia | 121 | 0.76 | 193 | 0.91 |
| Syria | 54 | 0.59 | 143 | 0.75 |
| Tunisia | 87 | 0.66 | 182 | 0.87 |
| Turkey | 120 | 0.76 | 202 | 0.95 |
| UAE | 94 | 0.68 | 186 | 0.88 |

The degree of a country represents the number of edges or countries it is connected to. The more a country has connections, the more influential it is in a network. All countries have a lower degree in mobility compared to collaboration as mentioned earlier.

Saudi Arabia has the highest degree in terms of mobility whereas Turkey has the highest value in the collaboration network. However, these two countries still top the MENA countries rankings in terms of degrees. When comparing the ranks of the degree for each country, some interesting values appear. Jordan and Morocco have the highest variation (-5) in terms of ranks of degrees compared to other countries. These two countries have relatively much more influence in the collaboration network than in the mobility network. Iran also exhibits a similar behaviour. On the other hand, Qatar has less influence in the collaboration than in mobility when benchmarked to other MENA countries. Pakistan and Oman show similar variations in terms of influence. Other countries have equivalent influences when degrees are ranked for each network

When analysing the closeness centrality, Turkey has also the highest closeness in Collaboration and Saudi Arabia has the highest value in mobility. The variations of closeness ranks are similar to the variations of degree ranks. The ranks of a given country in terms of degree and closeness in each network are of the same levels of the rank of this specific country in terms of number of scholars and publications. The MENA networks exhibit preferential connectivity or preferential attachment to specific countries (Barabási & Albert, 1999) such as Turkey, Saudi Arabia, Iran or Egypt. These countries or regional hubs play important roles in network development. In addition to North American and European countries, leading research countries in MENA tend to attract more researchers in terms of collaborative papers and mobility flows.

Finally, for each country in MENA, we distinguish two types of relations in the mobility and collaboration network: *MENA-MENA relations* and *Non-MENA relations*. Then, we compared the shares of MENA-MENA with the Non-MENA relations for the mobility and the collaboration phenomena for each individual country. Figure 9 shows the shares of collaboration and mobility relations by type and by country in MENA between 2008 and 2017. In general, both collaborations and mobility exhibit a stronger international than regional focus from a MENA perspective. From a country point of view, few cases such as Egypt or Saudi Arabia have a higher share of mobility exchanges with other MENA than with Non-MENA countries. To a lesser extent, Jordan and Kuwait also have a slightly higher share of MENA-MENA than Non-MENA mobility-exchanges. On the other hand, Iran, Turkey, Morocco, Algeria and Tunisia have a relatively low share (12.5%) of their papers with an author from another MENA country. These 5 countries show an average of 15% of mobility relations with the MENA region.

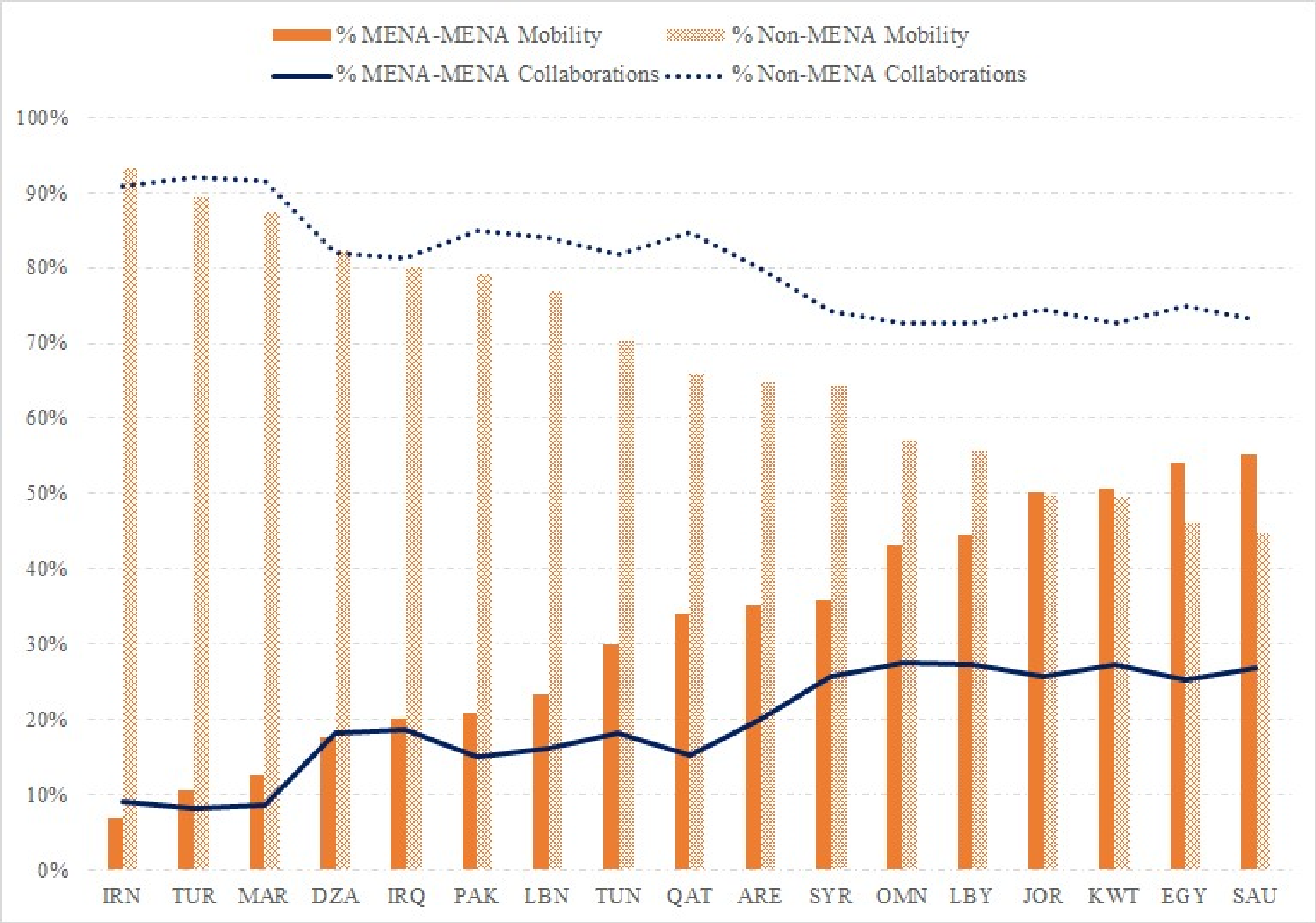

**Figure 9. Shares of %MENA-MENA collaboration and mobility relations by country in MENA ordered by percentage of MENA-MENA mobility ties (2008-2017).**

We also notice, for most countries in MENA, the shares of MENA-MENA mobility relations are higher than the shares of MENA-MENA collaboration relations. From the MENA region perspective, this suggests that the countries mobility links for these countries are more locally focused than the collaborations.

## 4. Discussion and conclusions

The main objective of this study was to better understand the scientific mobility flows in the Middle East and North Africa region. We extended previous research on macro-level indicators studies of scientific mobility using bibliometric indicators. Several results of our study confirm Scott's 'Fluid Globalization' framework (2015) where mobility is described as a 'spectrum', from the deeply rooted to the highly mobile scientists, with most scholars standing in the middle of that spectrum. The scientific mobility is a phenomenon within a wider context. The globalisation of the economy, proximities (geographical, social, cultural, linguistic and socio-political), the democratisation of mobility as well as the internationalisation all influence the scientific mobility. Some results also illustrate the 'Hegemonic internationalisation' framework (Scott, 2015). We observe large flows from/to Western Europe and the United States. Some mobility linkages suggest also an 'evolving core' including East Asian countries. These two frameworks offer interesting aspects that we illustrate in our study. However, they still focus on a single major 'core' and 'periphery' system. Leading research countries in MENA also tend to attract researchers in terms of mobility flows. Indeed, the common cultural spaces make the international mobility easier for scholars. Although Scott (2015) qualifies this type of scientific mobility as not 'remarkable', it is at least as important as mobility from the 'periphery' to the 'core'. Scientific mobility is often perceived as 'brain-drain' with flows from Non-West to West countries. This also applies to MENA. 'Brain-drain' is mainly used to describe the flows from MENA to non-MENA countries, especially Western countries (US and Europe). This study allows us to understand mobility from a local perspective. Similar claims have been made in other fields: a single core-periphery system is not efficient in cultural flows (Appadurai, 1996).

We now discuss in detail the main findings identified in our analysis. The country profiles as well as the demographic data of migrant scholars are informative for policy makers interested in the MENA region. In MENA, collaboration and mobility are quite aligned, although mobility in MENA is larger as compared to other studies (Chinchilla-Rodríguez et al., 2018). 12% of identified researchers have shown signs of international mobility. The *mobile* scientists are mainly *Directional Travellers* who represent 5.6% of the scholars of our dataset. *Migrant* is the second most common mobility type (3.2%). These shares illustrate the spectrum used by Scott (2015) to think about scientific mobility.

In this study, several characteristic patterns of the MENA region regarding the circulation of scholars can be highlighted. The MENA region is highly connected with Europe based on the number of mobile scientists. Europe is the first mobility destination and origin with 37% of the flows from/to MENA, followed by North America (24%), MENA (20%) and Asia (16%). Oceania, Africa, and South America show a much lower circulation of scholars (less than 3%). In terms of international destinations, the MENA region has a relatively high level of intra-regional mobility flows.

- Qatar, Saudi Arabia, United Arab Emirates, Kuwait can be described as *attracting countries*.
- Turkey, Egypt, Pakistan, Morocco, Algeria, Jordan and Lebanon are more *balanced countries*.
- Iran, Tunisia, Iraq and Syria can be considered as *sending countries*.

The region is highly connected with Europe based on the mobility flows of scientists. Europe is indeed the first mobility destination and origin, followed closely by North America. Asia is the third preferred destination and origin. Oceania, Africa and South America show a much lower circulation of scholars from and to MENA. At the country level, United States, France, United Kingdom, Germany, Canada, China, Malaysia, Italy, Japan and Australia are the main non-MENA destinations and origins. We retrieve here most Western countries mentioned by Scott (2015) with China and Malaysia from the Far-East. Some cases stand out when we look at specific MENA countries. Geographical, cultural, historical, linguistic and socio-political proximities have an influence on the mobility ties. For example, this is the case for France which is the preferred destination for scholars originating from its former colonies in MENA, specifically Morocco, Algeria and Tunisia. We also observe strong ties between North African countries with other countries in Europe such as Spain, Germany, Switzerland and Netherlands. United Kingdom appears to be one of the preferred destinations for scientists from GCC countries such as Saudi

Arabia, the United Arab Emirates and Qatar. Scholars from Egypt and Jordan have mainly migrated to Saudi Arabia, ahead of the United States. The observed flows confirm the geo-political considerations mentioned by Scott (2015): Attraction of ex-colonial powers or countries which speak 'world' languages, common cultural spaces, the key role of economic conditions, the 'big country small-country' effect, and political changes such as revolutions or civil unrest. Immigration restrictions, sanctions and travel bans affect mobility linkages such as in the case of Iran (Kokabisaghi et al 2019). Except for Pakistan and Iran, we can already find some countries outside of the region within the top 15 destinations/origins of MENA migrant scholars. Researchers from Pakistan migrate mainly from and to China. Iraq and, to a lesser extent, Iran have major flows from and to Malaysia. A previous study mentions, one in seven international students in Malaysia was of Iranian origin in 2012 (Unesco, 2015). Malaysia is one of the rare countries which do not impose visas on Iranian citizens.

The socio-political environment, cooperation and exchange programs could also contribute to explain some of the observed mobility flows. For example, the Pakistani Prime Minister Nawaz Sharif referred to Pakistan and China as *Iron Brothers* when the two countries signed in 2015 the China-Pakistan Economic Corridor (CPEC) (Vandewalle, 2015). The CPEC projects play an important role in China's *One Belt One Road* initiative. Later in 2017, China and Pakistan agreed to strengthen existing cooperation in Science and Technology. Europe and Mediterranean countries have also signed several bilateral research and innovation cooperation agreements such as Tunisia (2004), Morocco (2005), Egypt (2008), Jordan (2010) and Algeria (2013) (European Commission, March 2019). As part of the *5+5 Dialogue*, 5 countries from the Arab Maghreb Union (Morocco, Algeria, Tunisia, Mauritania, Libya) and 5 countries from the Western Mediterranean (Spain, Malta, Portugal, Italy, France) have regularly met since 1990 to discuss a wide range of issues (security, economic co-operation, defence, migration, education and renewable energy) (Unesco, 2015). In September 2013, the meeting focused on research and innovation and Ministers of scientific research from these countries signed the *Rabat Declaration (Rabat Declaration, 2013).* The ministers undertook the task to facilitate the scientific mobility by granting scientific-researcher visas, to promote the training of researchers, and to promote the transfer of technology and access to the scientific infrastructure.

From a demographic point of view, almost all MENA countries are dominated by male researchers. Countries such as Saudi Arabia, Iran, Jordan, the United Arab Emirates and Qatar have shown high degrees of male dominance (Larivière et al., 2013). We notice Pakistan and Iraq also have a high gender ratio. Tunisia, Lebanon, and Turkey are the only MENA countries for which the male-to-female ratio is close to 1. Compared to a developed country like Germany, these findings are in contrast with some of the previously published results by Zhao et al. (2021). In terms of mobility, mobile scholars in MENA are mainly men with a relatively senior academic status. These specificities are exacerbated in few countries such as Saudi Arabia, Iraq, Syria and Libya. Although GCC countries have a strong attraction of scholars, they seem to attract almost exclusively male researchers. There is a clear gender gap in terms of scientific mobility. Men represent 66% of all migrants in MENA. Women account for 12%. For the remaining authors, the gender was not identified reliably. We notice similar shares when comparing the emigrants and immigrants.

However, these shares vary by country. Tunisia and Lebanon have the highest shares of female emigrants, 22% and 21% respectively. These two countries are followed by Turkey, Algeria, Morocco, and Iran with around 17% of female migrant scholars. Egypt and Pakistan have a share of around 11% of female migrant scholars. In the remaining countries, women account for less than 10% of migrant scientists with the lowest shares in Iraq, Saudi Arabia, Syria, and Libya (about 7%). In all MENA countries, the gender disparity is more severe among the migrant researchers. The gender ratio among migrant scholars is on average 2.5 times higher than the gender ratio for all researchers. Our analysis allows us to explore the extent of the gender gaps in the MENA region and to understand how these disparities vary between migrants and all researchers by MENA country. The MENA countries have seen an increased participation of women in higher education, particularly in the GCC countries with 62 percent of enrolled students are female (Jaramillo et al., 2011). Although mobility is a means to opportunity (Hanson, 2010) by providing access to people, networks, and infrastructures that make their research more visible to influential researchers in their fields (Laudel, 2005), women are also more likely to bear responsibilities for children and households (Ackers, 2008; Xie et al., 2003). Metcalfe (2008) has shown that there is much to be gained by policies in the Middle East. Zippel (2011) has argued that national funding policies toward international mobility of scientists have gendered implications. Policy makers should adopt policies that support family burdens on women which would help them in their careers and would result in a more balanced research ecosystem (Karam & Afiouni, 2014). Such policies could include policies and practices of more flexible and temporary mobility as suggested by Cañibano et al. (2016).

The average academic age of migrant scholars was 12.39 years in MENA between 2008 and 2017. At the regional level, Emigrants have an average of academic age of 12.3 years versus 12.5 for the Immigrants. The academic age group '6 – 10' years is the most common for the immigrant and emigrant scholars and represent around 42% of all the migrants. '11 – 15' is the second age group, representing 32% of the migrant scientists. Migrant scholars with an academic age between 16 and 20 years represent a share of 10% of all the migrant authors. Other age groups had a share of less than 6%. As shown in Appendix B, the size of academic age group also varies by country. During the so called 'Arab Spring', young citizens have clearly asked for more and better development opportunities. The MENA countries stand at different levels of economic development, but they all share an interest in the higher education supply and demand. From an internationalization perspective, policies have implications on these three areas that have been discussed by a World Bank group of authors for the MENA region (Jaramillo et al., 2011). The same authors have mentioned the importance to look at the policy framework to improve the quality and relevance of higher education systems in the MENA region. For example, they note that a key driver for internationalization is demographic trends. MENA countries have large young populations and increasing numbers of students. To meet such high demand, cross-border higher education is widely used by developing joint research and development programs. Traditional university partnerships, probably the most common form of international mobility of higher education, also contribute to mobility flows of PhD students, post-docs, and relatively more senior researchers.

In general, both collaborations and mobility show a stronger international than regional focus from the MENA region perspective. We note the role of United States and United Kingdom as important actors driving collaboration with most of MENA countries. Saudi Arabia, Iran, Egypt, and Turkey driving most of the international cooperation within the region. However, their partnerships seem to vary. While Iran, Egypt and Saudi Arabia have strong collaboration ties with Asian countries, Turkey's main collaborating countries include several European countries such as Germany and France.

From a country point of view, few cases such as Egypt or Saudi Arabia have a higher share of mobility exchanges with other MENA than with Non-MENA countries. Similarly, but to a lesser extent, Jordan and Kuwait have a slightly higher share of MENA-MENA than Non-MENA mobility-exchanges. On the other hand, Iran, Turkey, Morocco, Algeria and Tunisia have a relatively low share (12.5%) of their papers with an author from another MENA country. For these 5 countries, the mobility relations with the MENA region represent 15% of all their mobility linkages. On this aspect, there have been some calls at the First Arab-Euro Conference to develop stronger and closer collaboration between Arab countries to have more Arab researchers returning to and more Europeans visiting the MENA region (Vesper, 2013). For most countries in MENA, the shares of MENA-MENA mobility relations are higher than the shares of MENA-MENA collaboration relations. From the MENA region perspective, this suggests that the countries mobility links for these countries are more locally focused than the collaborations.

In terms of methodology, this study represents a blueprint of how scientometric studies can inform the mobility dynamics of specific countries and geographical regions. We acknowledge that future studies are still necessary to more discuss the validity and reliability of scientometric approaches to capture scientific mobility and its diversity. In Sugimoto et al (2017) there was already some discussions regarding the contextualization of scientometric mobility data by comparing it to ORCID data, and this is an approach that will need more attention. However, the use of ORCID to validate scientific mobility, although relevant, is also not free of its own limitations. For example, ORCID, but arguably also other type of mobility survey data, suffer from limitations of coverage, representativeness, lack of standardization or completeness (Gomez et al., 2020). This means that currently there is no established "golden set" to determine scientific mobility flows at the global level. Therefore, the use of scientometric data to study scientific mobility must be seen as an informative but conservative approach, needing to observe the intrinsic limitations coming from the method (cf. Robinson-Garcia et al, 2019), and whenever possible be used in combination with other sources of mobility information. In line with this, in this paper we intend to provide useful material for the analysis and discussion of scientific mobility in the MENA region as well as statistical information to issues raised already since the early 2000s by the Observatory of International Migration in the Arab Region in collaboration with the United Nations (2002-2018). We also complemented previous studies where data was limited to OECD countries as destinations of scientists (Fargues, 2006; Özden, 2006). Future research should focus on expanding these analytical capabilities in order to study other geographical areas (e.g. South America, Sub-Saharan Africa, Sahel region, OECD countries, before and after Brexit effects). Such analyses will be necessary to better support the assessment of different scientific systems, and to determine how geopolitical decisions have impact on the collaboration and circulation of researchers and scientific ideas. The approach we used to measure mobility relies on tracking the change of the author affiliation at the country level. We acknowledge that the taxonomy of mobility used in our study is not absolute. Not every change of affiliation should be interpreted as an indicator of breaking ties with the original countries of the researchers, particularly in the case of the travellers, who have multiple affiliation over time. Other classes could also be introduced and discussed. There are many different types of mobility that could be derived such as *return migrants* or *transients* as defined and used in other studies (Moed & Halevi, 2014; Subbotin & Aref, 2020). Future research may seek to use the approach presented by Sugimoto et al. (2016) to estimate the mobility at the regional, city and institutional levels in MENA, as well as including other typologies of mobility flows such as the return of mobile researchers, as well as the more transient type of mobility relationships (i.e. researchers with just an occasional – one time – affiliation relationship with a country, cf. Moed and Halevi (2014)). This granularity will enable us to capture the more domestic scholarly movements, as well as the role of those researchers who in some way return to their countries of origin. Such developments would substantially contribute to better inform the phenomenon of scientific mobility by also incorporating more local and dynamic perspectives. We also plan to combine the mobility indicators with other bibliometric information such as citation metrics, research areas or funding acknowledgments. The further improvement and development of advanced scientometric mobility studies will also benefit decision-makers and science policy analysts who look for programs and strategies that will encourage international collaborations and mobility (e.g., China Scholarship Council, Marie Sklodowska-Curie or Ramón y Cajal fellowships programs).

## Author contributions

JEO: conceived the study, participated in its design and coordination, gathered data, generated figures, interpreted the data, and wrote the paper;
NRG: participated in the design and coordination of the study, interpreted the data, and wrote the paper;
RC: participated in the design and coordination of the study, interpreted the data, and wrote the paper.

## Acknowledgments

We would like to thank Ludo Waltman and Thomas Franssen for providing valuable comments on an earlier version of the manuscript. We are also grateful for the feedback from two reviewers. RC was partially funded by the South African DST-NRF Center of Excellence in Scientometrics and Science, Technology, and Innovation Policy (SciSTIP).

## References

Abramo, G., D'Angelo, C. A., & Solazzi, M. (2011). The relationship between scientists' research performance and the degree of internationalization of their research. *Scientometrics, 86*(3), 629-643. doi:10.1007/s11192-010-0284-7

Ackers, L. (2005). Moving people and knowledge: Scientific mobility in the European Union. *International Migration, 43*(5), 99-131. doi:10.1111/j.1468-2435.2005.00343.x

Ackers, L. (2008). Internationalisation, Mobility and Metrics: A New Form of Indirect Discrimination? *Minerva, 46*(4), 411-435. doi:10.1007/s11024-008-9110-2

Altbach, P. G., & Knight, J. (2007). The Internationalization of Higher Education: Motivations and Realities. *Journal of Studies in International Education, 11*(3-4), 290-305. doi:10.1177/1028315307303542

Appadurai, A. (1996). *Modernity at large : cultural dimensions of globalization*: Minneapolis, Minn: University of Minnesota Press.

Aref, S., Zagheni, E., & West, J. (2019). The Demography of the Peripatetic Researcher: Evidence on Highly Mobile Scholars from the Web of Science *Lecture Notes in Computer Science* (Vol. 11864, pp. 50-65): Springer International Publishing.

Backes, T. (2018, 2018). *Effective Unsupervised Author Disambiguation with Relative Frequencies*.

Baldwin, G. B. (1970). Brain drain or overflow? *Foreign Affairs, 48*(2), 358-372.

Barabási, A.-L., & Albert, R. (1999). Emergence of scaling in random networks. *Science, 286*(5439), 509-512.

Barabási, A. L., Jeong, H., Néda, Z., Ravasz, E., Schubert, A., & Vicsek, T. (2002). Evolution of the social network of scientific collaborations. *Physica A: Statistical Mechanics and its Applications, 311*(3-4), 590-614. doi:10.1016/s0378-4371(02)00736-7

Beine, M., Docquier, F., & Rapoport, H. (2008). Brain drain and human capital formation in developing countries: Winners and losers. *Economic Journal, 118*(528), 631-652. doi:10.1111/j.1468-0297.2008.02135.x

Bernard, M., Bernela, B., & Ferru, M. (2021). Does the geographical mobility of scientists shape their collaboration network? A panel approach of chemists' careers. *Papers in Regional Science, 100*(1). doi:10.1111/pirs.12563

Boulding, K. E. (1966). ECONOMICS OF KNOWLEDGE AND KNOWLEDGE OF ECONOMICS. *American Economic Review, 56*(2), 1-13.

Boulding, K. E. (1966). The economics of knowledge and the knowledge of economics. *The American Economic Review, 56*(1/2), 1-13.

Breschi, S., & Lissoni, F. (2009). Mobility of skilled workers and co-invention networks: an anatomy of localized knowledge flows. *Journal of Economic Geography, 9*(4), 439-468. doi:10.1093/jeg/lbp008

Cañibano, C., Fox, M. F., & Otamendi, F. J. (2016). Gender and patterns of temporary mobility among researchers. *Science and Public Policy, 43*(3), 320-331. doi:10.1093/scipol/scv042

Cañibano, C., & Woolley, R. (2015). Towards a Socio-Economics of the Brain Drain and Distributed Human Capital. *International Migration, 53*(1), 115-130. doi:10.1111/imig.12020

Caron, E., & van Eck, N. J. (2014). *Large scale author name disambiguation using rule-based scoring and clustering.* Paper presented at the Proceedings of the 19th international conference on science and technology indicators.

Cavacini, A. (2016). Recent trends in Middle Eastern scientific production. *Scientometrics, 109*(1), 423-432. doi:10.1007/s11192-016-1932-3

Chinchilla-Rodríguez, Z., Miao, L., Murray, D., Robinson-García, N., Costas, R., & Sugimoto, C. R. (2018). A Global Comparison of Scientific Mobility and Collaboration According to National Scientific Capacities. *Frontiers in Research Metrics and Analytics, 3*. doi:10.3389/frma.2018.00017

Cota, R. G., Gonçalves, M. A., & Laender, A. H. (2007). *A Heuristic-based Hierarchical Clustering Method for Author Name Disambiguation in Digital Libraries.* Paper presented at the SBBD.

D'Angelo, C. A., & Van Eck, N. J. (2020). Collecting large-scale publication data at the level of individual researchers: a practical proposal for author name disambiguation. *Scientometrics, 123*(2), 883-907. doi:10.1007/s11192-020-03410-y

De Nooy, W., Mrvar, A., & Batagelj, V. (2018). *Exploratory social network analysis with Pajek: Revised and expanded edition for updated software* (Vol. 46): Cambridge University Press.

Di Maria, C., & Stryszowski, P. (2009). Migration, human capital accumulation and economic development. *90*(2), 306-313. doi:10.1016/j.jdeveco.2008.06.008

Docquier, F., & Marfouk, A. (2005). International Migration by Educational Attainment 1990–2000 (Release 1).

Dokko, G., & Rosenkopf, L. (2010). Social Capital for Hire? Mobility of Technical Professionals and Firm Influence in Wireless Standards Committees. *Organization Science, 21*(3), 677-695. doi:10.1287/orsc.1090.0470

Dokko, G., Wilk, S. L., & Rothbard, N. P. (2009). Unpacking prior experience: How career history affects job performance. *Organization Science, 20*(1), 51-68. doi:10.1287/orsc.1080.0357

European Commission. (March 2019). Mediterranean Partners. Retrieved from https://ec.europa.eu/research/iscp/index.cfm?pg=med_part

Fargues, P. (2006). *International migration in the Arab region: Trends and policies.* Paper presented at the United Nations Expert Group Meeting on International Migration and Development in the Arab Region, Beirut.

Franceschet, M., & Costantini, A. (2010). The effect of scholar collaboration on impact and quality of academic papers. *Journal of Informetrics, 4*(4), 540-553. doi:10.1016/j.joi.2010.06.003

Gazni, A., Sugimoto, C. R., & Didegah, F. (2012). Mapping World Scientific Collaboration: Authors, Institutions, and Countries. *Journal of the American Society for Information Science and Technology, 63*(2), 323-335. doi:10.1002/asi.21688

Glanzel, W. (2001). National characteristics in internationalscientific co-authorship relations. *Scientometrics, 51*(1), 69-115. doi:10.1023/a:1010512628145

Glytsos, N. P. (2010). Theoretical considerations and empirical evidence on brain drain grounding the review of Albania's and Bulgaria's experience1. *International Migration, 48*(3), 107-130.

Gomez, C. J., Herman, A. C., & Parigi, P. (2020). Moving more, but closer: Mapping the growing regionalization of global scientific mobility using ORCID. *Journal of Informetrics, 14*(3), 101044. doi:10.1016/j.joi.2020.101044

Gul, S., Nisa, N. T., Shah, T. A., Gupta, S., Jan, A., & Ahmad, S. (2015). Middle East: research productivity and performance across nations. *Scientometrics, 105*(2), 1157-1166. doi:10.1007/s11192-015-1722-3

Hanneman, R. A., & Riddle, M. (2005). Introduction to social network methods: University of California Riverside.

Hanson, S. (2010). Gender and mobility: new approaches for informing sustainability. *Gender, Place & Culture, 17*(1), 5-23. doi:10.1080/09663690903498225

Hassan Al Marzouqi, A. H., Alameddine, M., Sharif, A., & Alsheikh-Ali, A. A. (2019). Research productivity in the United Arab Emirates: A 20-year bibliometric analysis. *Heliyon, 5*(12), e02819. doi:10.1016/j.heliyon.2019.e02819

Hayek, F. A. (1945). The Use of Knowledge in Society. *American Economic Review, 35*(4), 519-530.

International Labour Office. (2009). *International labour migration and employment in the Arab region: Origins, consequences and the way forward*. Retrieved from

Jaffe, A. B., Trajtenberg, M., & Henderson, R. (1993). Geographic Localization of Knowledge Spillovers as Evidenced by Patent Citations. *The quarterly journal of economics, 108*(3), 577-598. doi:10.2307/2118401

Jaramillo, A., Ruby, A., Henard, F., & Zaafrane, H. (2011). Internationalization of Higher Education in MENA: Policy issues associated with skills formation and mobility.

Johnson, H. G. (1965). The Economics of the Brain Drain - The Canadian Case. *Minerva, 3*(3), 299-311. doi:10.1007/bf01099956

Karam, C. M., & Afiouni, F. (2014). Localizing women's experiences in academia: multilevel factors at play in the Arab Middle East and North Africa. *The International Journal of Human Resource Management, 25*(4), 500-538. doi:10.1080/09585192.2013.792857

Kato, M., & Ando, A. (2017). National ties of international scientific collaboration and researcher mobility found in Nature and Science. *Scientometrics, 110*(2), 673-694. doi:10.1007/s11192-016-2183-z

Kidd, C. V. (1965). The Economics of the Brain-Drain. *Minerva, 4*(1), 105-107. doi:10.1007/bf01585988

Kokabisaghi, F., Miller, A. C., Bashar, F. R., Salesi, M., Zarchi, A. A. K., Keramatfar, A., . . . Vahedian-Azimi, A. (2019). Impact of United States political sanctions on international collaborations and research in Iran. *BMJ Global Health, 4*(5), e001692. doi:10.1136/bmjgh-2019-001692

Krishna, V., & Khadria, B. (1997). Phasing scientific migration in the context of brain gain and brain drain in India. *Science, Technology and Society, 2*(2), 347-385.

Larivière, V., Ni, C., Gingras, Y., Cronin, B., & Sugimoto, C. R. (2013). Bibliometrics: Global gender disparities in science. *Nature News, 504*(7479), 211.

Laudel, G. (2003). Studying the brain drain: Can bibliometric methods help? *Scientometrics, 57*(2), 215-237. doi:10.1023/a:1024137718393

Laudel, G. (2005). Migration Currents Among the Scientific Elite. *Minerva, 43*(4), 377-395. doi:10.1007/s11024-005-2474-7

League of Arab States. (2009). *Regional Report on Arab Labour Migration: Brain Drain or Brain Gain?* Retrieved from Cairo, Egypt: League of Arab States.:

Levin, S. G., & Stephan, P. E. (1999). Are the foreign born a source of strength for US science? : American Association for the Advancement of Science.

Lowell, B. L. (2003). The need for policies that meet the needs of all. *Science and Development Network.*

Malakhov, V. A., & Erkina, D. S. (2020). Russian Mathematicians in the International Circulation of Scientific Personnel: Bibliometric Analysis. *Sociologia Nauki I Tehnologij-Sociology of Science & Technology, 11*(1), 63-74. doi:10.24411/2079-0910-2020-11005

Mawdsley, J. K., & Somaya, D. (2016). Employee Mobility and Organizational Outcomes: An Integrative Conceptual Framework and Research Agenda. *42*(1), 85-113. doi:10.1177/0149206315616459

Metcalfe, B. D. (2008). Women, Management and Globalization in the Middle East. *Journal of Business Ethics, 83*(1), 85-100. doi:10.1007/s10551-007-9654-3

Meyer, J. B. (2001). Network Approach versus Brain Drain: Lessons from the Diaspora. *International Migration, 39*(5), 91-110. doi:10.1111/1468-2435.00173
Miguélez, E., & Moreno, R. (2013). Research Networks and Inventors' Mobility as Drivers of Innovation: Evidence from Europe. *Regional Studies, 47*(10), 1668-1685. doi:10.1080/00343404.2011.618803
Mihaljević, H., & Santamaría, L. (2021). Disambiguation of author entities in ADS using supervised learning and graph theory methods. *Scientometrics, 126*(5), 3893-3917. doi:10.1007/s11192-021-03951-w
Miranda-González, A., Aref, S., Theile, T., & Zagheni, E. (2020). Scholarly migration within Mexico: analyzing internal migration among researchers using Scopus longitudinal bibliometric data. *EPJ Data Science, 9*(1), 34. doi:10.1140/epjds/s13688-020-00252-9
Moed, H. F., Aisati, M. h., & Plume, A. (2012). Studying scientific migration in Scopus. *Scientometrics, 94*(3), 929-942. doi:10.1007/s11192-012-0783-9
Moed, H. F., & Halevi, G. (2014). A bibliometric approach to tracking international scientific migration. *101*(3), 1987-2001. doi:10.1007/s11192-014-1307-6
Morano Foadi, S. (2006). Key issues and Causes of the Italian Brian Drain. *Innovation: The European Journal of Social Science Research, 19*(2), 209-223. doi:10.1080/13511610600804315
Mountford, A. (1997). Can a brain drain be good for growth in the source economy? *Journal of Development Economics, 53*(2), 287-303. doi:10.1016/s0304-3878(97)00021-7
Nane, G. F., Larivière, V., & Costas, R. (2017). Predicting the age of researchers using bibliometric data. *Journal of Informetrics, 11*(3), 713-729. doi:10.1016/j.joi.2017.05.002
Nerad, M. (2010). Globalization and the internationalization of graduate education: A macro and micro view. *Canadian Journal of Higher Education, 40*(1), 1-12.
Netz, N., Hampel, S., & Aman, V. (2020). What effects does international mobility have on scientists' careers? A systematic review. *Research Evaluation, 29*(3), 327-351. doi:10.1093/reseval/rvaa007
Newman, M. E. J. (2002). Assortative Mixing in Networks. *Physical Review Letters, 89*(20). doi:10.1103/physrevlett.89.208701
OECD. (2008). The Global Competition for Talent: Mobility of the Highly Skilled. Retrieved from www.oecd.org/sti/inno/theglobalcompetitionfortalentmobilityofthehighlyskilled
OECD. (2015). *Connecting with Emigrants - A Global Profile of Diasporas 2015*. Retrieved from https://www.oecd-ilibrary.org/content/publication/9789264239845-en
Oppenheimer, J. R. (1948). The eternal apprentice. *Time magazine, 52,* 70-81.
Özden, Ç. (2006). Brain drain in Middle East and North Africa—The patterns under the surface. *United Nations Population, EGM/2006/10, New York, NY*.
Palomeras, N., & Melero, E. (2010). Markets for Inventors: Learning-by-Hiring as a Driver of Mobility. *Management Science, 56*(5), 881-895. doi:10.1287/mnsc.1090.1135
Rabat Declaration. (2013). Retrieved from https://pin.enssup.gov.ma/index.php/cooperation/cooperation-regionale/2-non-categorise/30-dialogue-5-5
Robinson-García, N., Cañibano, C., Woolley, R., & Costas, R. (2016). *Scientific mobility of early career researchers in Spain and The Netherlands through their publications.* Paper presented at the 21st International Conference on Science and Technology Indicators-STI 2016. Book of Proceedings.

Robinson-García, N., Sugimoto, C. R., Murray, D., Yegros-Yegros, A., Larivière, V., & Costas, R. (2018). Scientific mobility indicators in practice: International mobility profiles at the country level. *arXiv preprint arXiv:1806.07815*.

Robinson-García, N., Sugimoto, C. R., Murray, D., Yegros-Yegros, A., Larivière, V., & Costas, R. (2019). The many faces of mobility: Using bibliometric data to measure the movement of scientists. *Journal of Informetrics, 13*(1), 50-63. doi:10.1016/j.joi.2018.11.002

Santamaría, L., & Mihaljević, H. (2018). Comparison and benchmark of name-to-gender inference services. *PeerJ Computer Science, 4*, e156. doi:10.7717/peerj-cs.156

Sarwar, R., & Hassan, S.-U. (2015). A bibliometric assessment of scientific productivity and international collaboration of the Islamic World in science and technology (S&T) areas. *Scientometrics, 105*(2), 1059-1077. doi:10.1007/s11192-015-1718-z

Scellato, G., Franzoni, C., & Stephan, P. (2015). Migrant scientists and international networks. *Research Policy, 44*(1), 108-120. doi:10.1016/j.respol.2014.07.014

Schmoch, U., Fardoun, H. M., & Mashat, A. S. (2016). Establishing a World-Class University in Saudi Arabia: intended and unintended effects. *Scientometrics, 109*(2), 1191-1207. doi:10.1007/s11192-016-2089-9

Schulz, C., Mazloumian, A., Petersen, A. M., Penner, O., & Helbing, D. (2014). Exploiting citation networks for large-scale author name disambiguation. *EPJ Data Science, 3*(1). doi:10.1140/epjds/s13688-014-0011-3

Scott, P. (2015). Dynamics of Academic Mobility: Hegemonic Internationalisation or Fluid Globalisation. *European Review, 23*(S1), S55-S69. doi:10.1017/s1062798714000775

Shin, J. C., Lee, S. J., & Kim, Y. (2012). Knowledge-based innovation and collaboration: a triple-helix approach in Saudi Arabia. *Scientometrics, 90*(1), 311-326. doi:10.1007/s11192-011-0518-3

Siddiqi, A., Stoppani, J., Anadon, L. D., & Narayanamurti, V. (2016). Scientific Wealth in Middle East and North Africa: Productivity, Indigeneity, and Specialty in 1981–2013. *PLoS One, 11*(11), e0164500. doi:10.1371/journal.pone.0164500

Singh, J. (2005). Collaborative Networks as Determinants of Knowledge Diffusion Patterns. *Management Science, 51*(5), 756-770. doi:10.1287/mnsc.1040.0349

Singh, J., & Agrawal, A. (2011). Recruiting for Ideas: How Firms Exploit the Prior Inventions of New Hires. *Management Science, 57*(1), 129-150. doi:10.1287/mnsc.1100.1253

Slavova, K., Fosfuri, A., & De Castro, J. O. (2015). Learning by Hiring: The Effects of Scientists' Inbound Mobility on Research Performance in Academia. *Organization Science*. doi:10.1287/orsc.2015.1026

Song, H. (1997). From brain drain to reverse brain drain: Three decades of Korean experience. *Science, Technology and Society, 2*(2), 317-345.

Sonnenwald, D. H. (2007). Scientific collaboration. *Annual Review of Information Science and Technology, 41*(1), 643-681. doi:10.1002/aris.2007.1440410121

Stephan, P. E., & Levin, S. G. (2001). Exceptional contributions to US science by the foreign-born and foreign-educated. *Population Research and Policy Review, 20*(1/2), 59-79. doi:10.1023/a:1010682017950

Subbotin, A., & Aref, S. (2020). Brain Drain and Brain Gain in Russia: Analyzing International Migration of Researchers by Discipline using Scopus Bibliometric Data 1996-2020. *arXiv pre-print server arxiv:2008.03129*.

Sugimoto, C. R., Robinson-García, N., & Costas, R. (2016). Towards a global scientific brain: Indicators of researcher mobility using co-affiliation data. *OECD STI*.

Sugimoto, C. R., Robinson-García, N., Murray, D. S., Yegros-Yegros, A., Costas, R., & Lariviere, V. (2017). Scientists have most impact when they're free to move. *Nature, 550*(7674), 29-31. doi:10.1038/550029a

Tekles, A., & Bornmann, L. (2020). Author name disambiguation of bibliometric data: A comparison of several unsupervised approaches. *Quantitative Science Studies, 1*(4), 1510-1528. doi:10.1162/qss_a_00081

Times Higher Education. (2017). Latin American science funding crisis fuels brain drain. Retrieved from https://www.timeshighereducation.com/news/latin-american-science-funding-crisis-fuels-brain-drain

Torvik, V. I., & Smalheiser, N. R. (2009). Author name disambiguation in MEDLINE. *ACM Transactions on Knowledge Discovery from Data (TKDD), 3*(3), 1-29.

Unesco. (2015). *UNESCO science report: towards 2030*. Retrieved from https://unesdoc.unesco.org/ark:/48223/pf0000235406

United Nations. (2002-2018). *Coordination Meeting on International Migration*. Retrieved from https://www.un.org/en/development/desa/population/migration/events/coordination/index.asp

van Eck, N. J., & Waltman, L. (2009). VOSviewer: A Computer Program for Bibliometric Mapping. In B. Larsen & J. Leta (Eds.), *Proceedings of Issi 2009 - 12th International Conference of the International Society for Scientometrics and Informetrics, Vol 2* (Vol. 2, pp. 886-897). Leuven: Int Soc Scientometrics & Informetrics-Issi.

Van Raan, A. F. J. (1998). The influence of international collaboration on the impact of research results. *Scientometrics, 42*(3), 423-428. doi:10.1007/bf02458380

Vandewalle, L. (2015). *Pakistan and China: 'Iron brothers' forever?* Retrieved from http://www.europarl.europa.eu/RegData/etudes/IDAN/2015/549052/EXPO_IDA%282015%29549052_EN.pdf

Vesper, I. (2013). Euro-Arab science collaboration in the spotlight. *Research Europe*.

Wang, J., Hooi, R., Li, A. X., & Chou, M.-H. (2019). Collaboration patterns of mobile academics: The impact of international mobility. *Science and Public Policy, 46*(3), 450-462. doi:10.1093/scipol/scy073

Wang, Y. Q., Luo, H., & Shi, Y. Y. (2019). Complex network analysis for international talent mobility based on bibliometrics. *International Journal of Innovation Science, 11*(3), 419-435. doi:10.1108/ijis-04-2019-0044

Wasserman, S., & Faust, K. (1994). *Social network analysis: Methods and applications* (Vol. 8): Cambridge university press.

Watts, D. J., & Strogatz, S. H. (1998). Collective dynamics of ‘small-world’ networks. *Nature, 393*(6684), 440-442. doi:10.1038/30918

Wilsdon, J. (2011). Knowledge, networks and nations: Global scientific collaboration in the 21st century. *London: The Royal Society*.

World Bank. (June 2019). World Development Indicators. Retrieved from https://datahelpdesk.worldbank.org/knowledgebase/articles/906519-world-bank-country-and-lending-groups

World Bank. (October 2019). Middle East and North Africa. Retrieved from https://www.worldbank.org/en/region/mena

Xie, Y., Shauman, K. A., & Shauman, K. A. (2003). *Women in science: Career processes and outcomes* (Vol. 26): Harvard university press Cambridge, MA.

Yurevich, M. A., Erkina, D. S., & Tsapenko, I. P. (2020). Measuring International Mobility Of Russian Scientists: A Bibliometric Approach. *Mirovaya Ekonomika I Mezhdunarodnye Otnosheniya, 64*(9), 53-62. doi:10.20542/0131-2227-2020-64-9-53-62

Zhao, X., Aref, S., Zagheni, E., & Stecklov, G. (2021). International Migration in Academia and Citation Performance: An Analysis of German-Affiliated Researchers by Gender and Discipline Using Scopus Publications 1996-2020. *arXiv pre-print server*. doi:arxiv:2104.12380

Zippel, K. (2011). How gender neutral are state policies on science and international mobility of academics? *Sociologica*(1), 0-0.

Zweig, D. (1997). To return or not to return? Politics vs. economics in China's brain drain. *Studies in Comparative International Development, 32*(1), 92-125. doi:10.1007/bf02696307

## Appendix

Appendix A represents the shares of female and male authors for each country in MENA between 2008 and 2017. It also includes the share of authors with an unknown gender. A male gender was inferred to 57% of the disambiguated authors affiliated to a MENA country during the study period. For 33% of them, a female gender was inferred while no gender could be inferred to the remaining 10%.

**Appendix A. Shares of authors per gender and per country (2008-2017).**

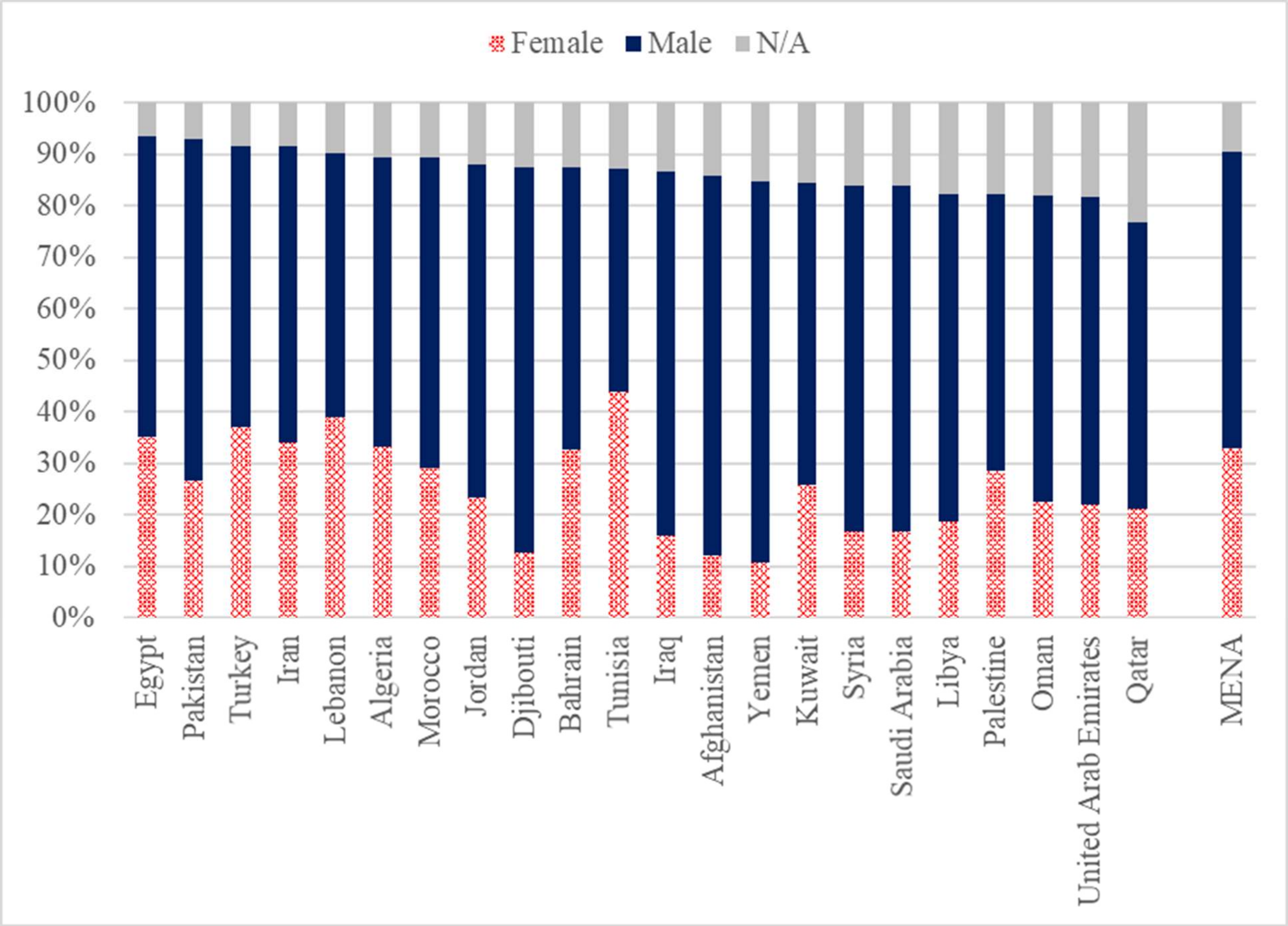

In Appendix B, we represent the origins and the destinations of mobile researchers in alluvial diagrams. Here, the diagrams focus only on emigrants and immigrants for countries where we have more than 1,000 mobile researchers. We constructed the alluvial diagrams for each country as follows. They include three steps:

- The first is Gender, with three nodes, *Male*, *Female* and *Not Available (N/A)*.
  The size of the nodes is proportional to the number of nodes containing that value.
- The second step is Academic Age group. Also, in this case the size of each node is proportional to the number of scholars with the average academic age within each 5 years range.
- The third is Country (of origin for Immigrants or destination for Emigrants).

The flows among nodes represent the number of scholars in our dataset sharing the combination of the three mentioned values: Gender-Academic Age-Country.

We also limited our analysis to the top 15 origins and destinations by number of migrant scholars for each country. The left charts represent the flows of scholar immigrants with their origins (*Immigrating from*).
The right charts show the flows of scholar emigrants along with their destinations (*Emigrating to*).

**Appendix B. Migration flows of scholars per gender and age group (2008-2017).**

Saudi Arabia:

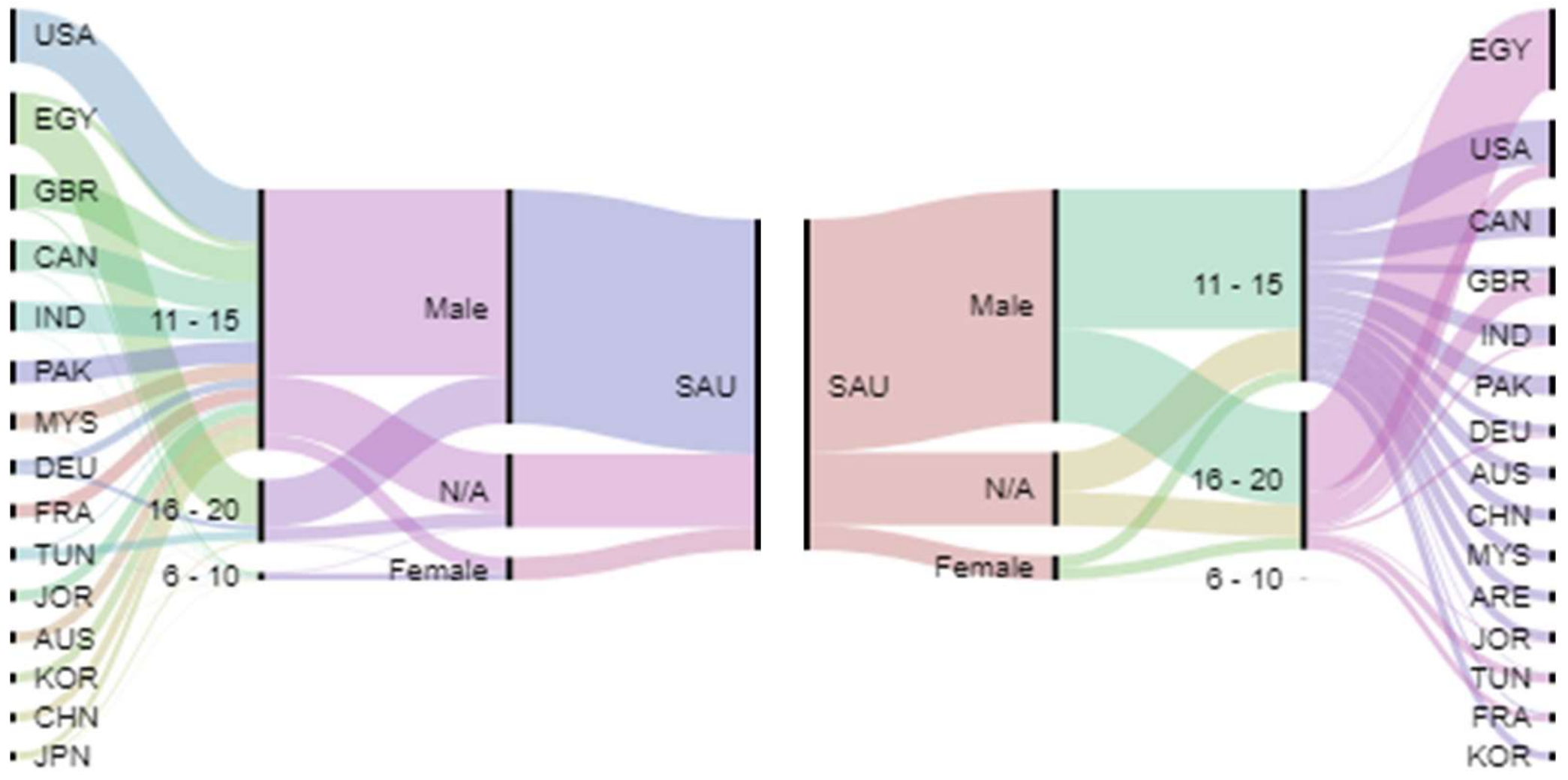

Egypt:

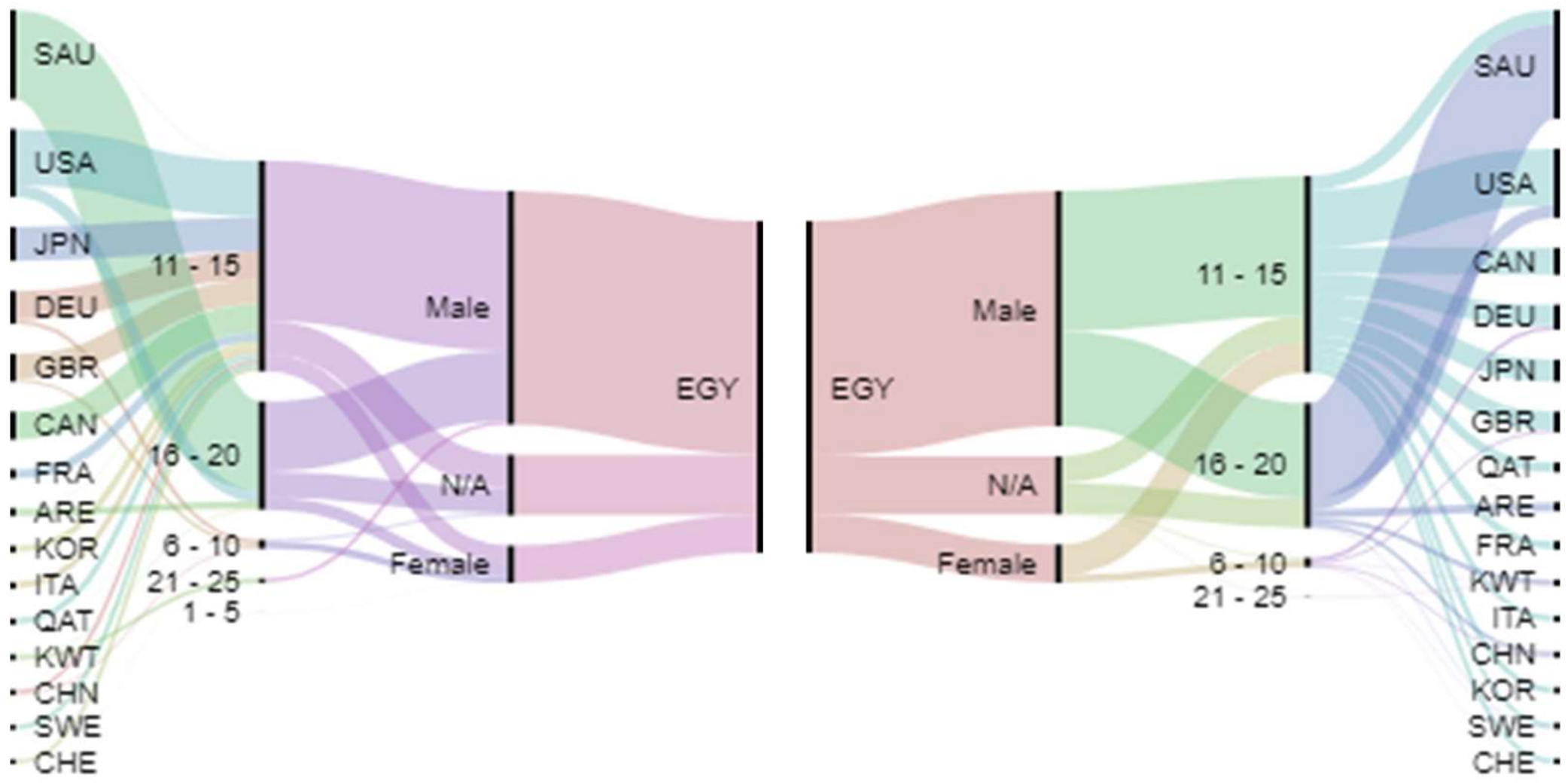

Turkey:

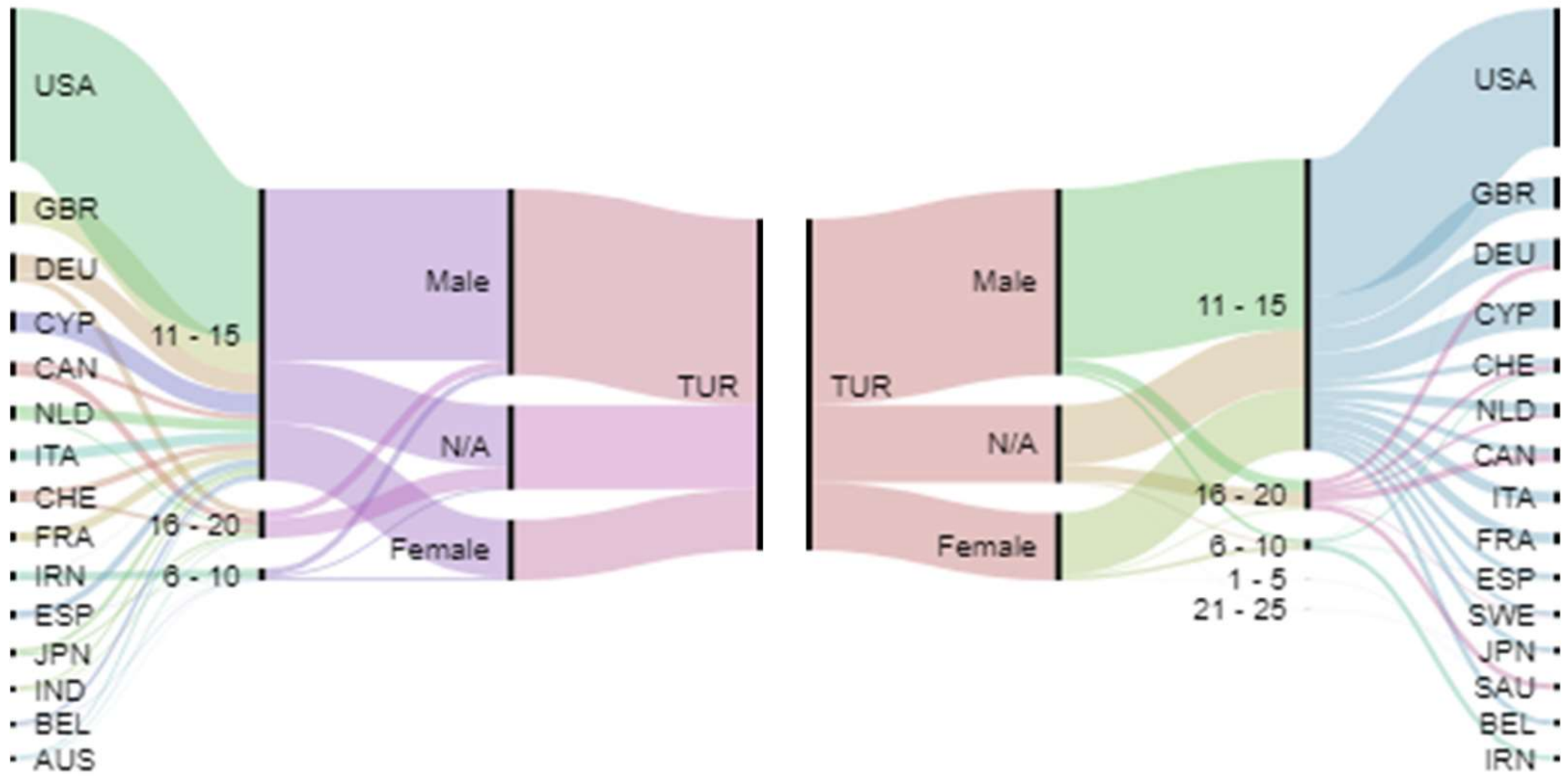

Iran:

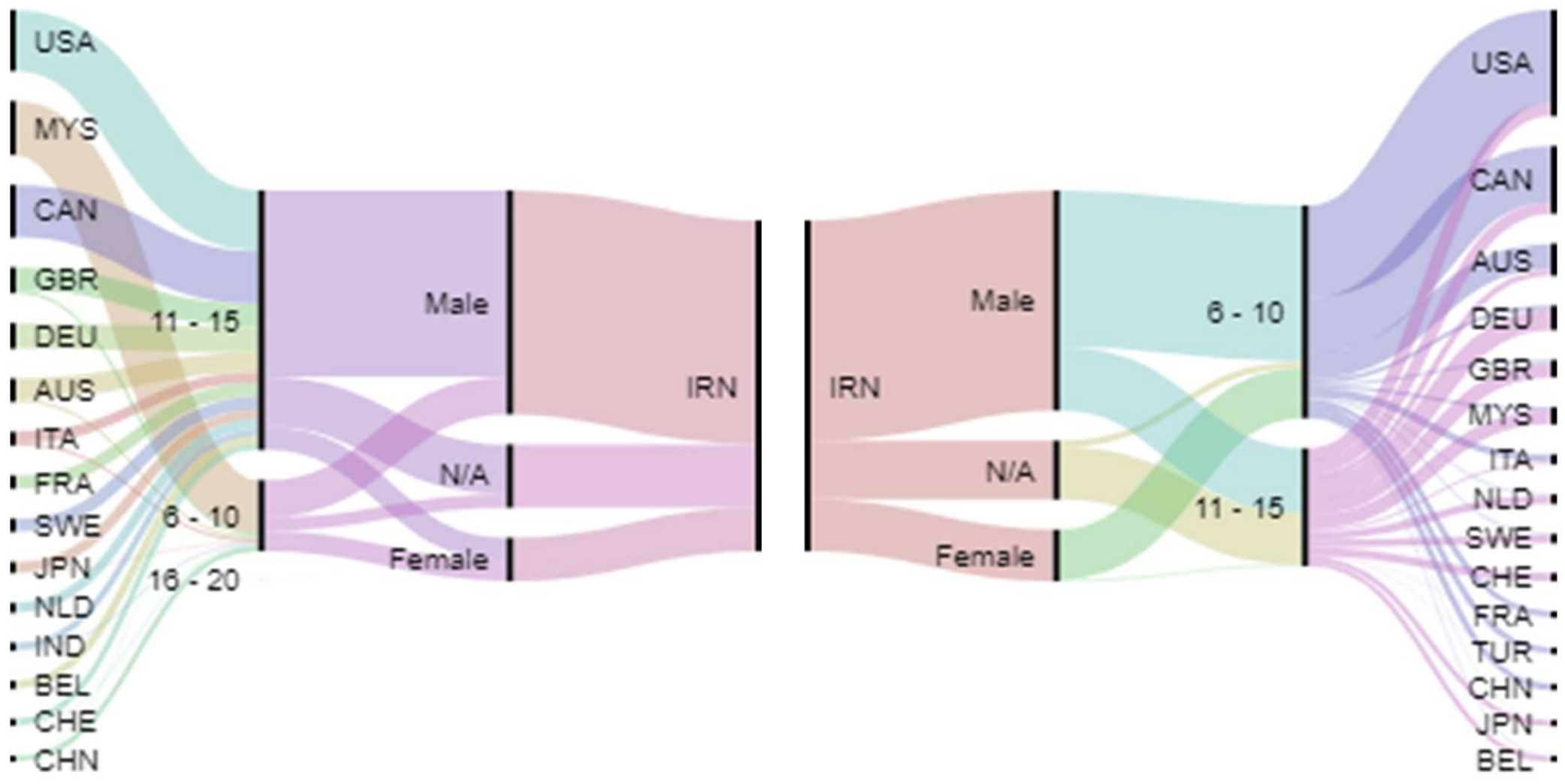

Pakistan:

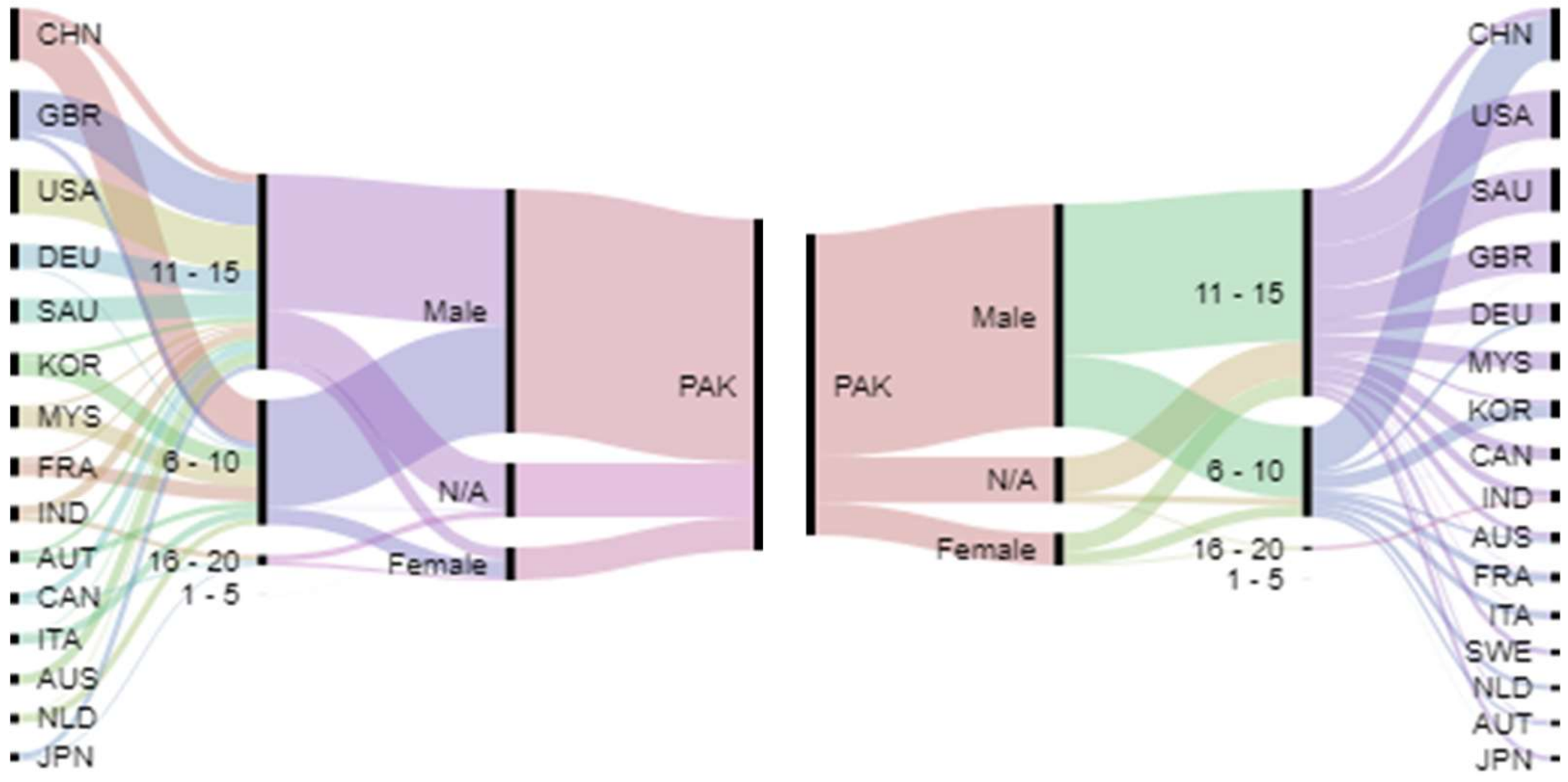

Tunisia:

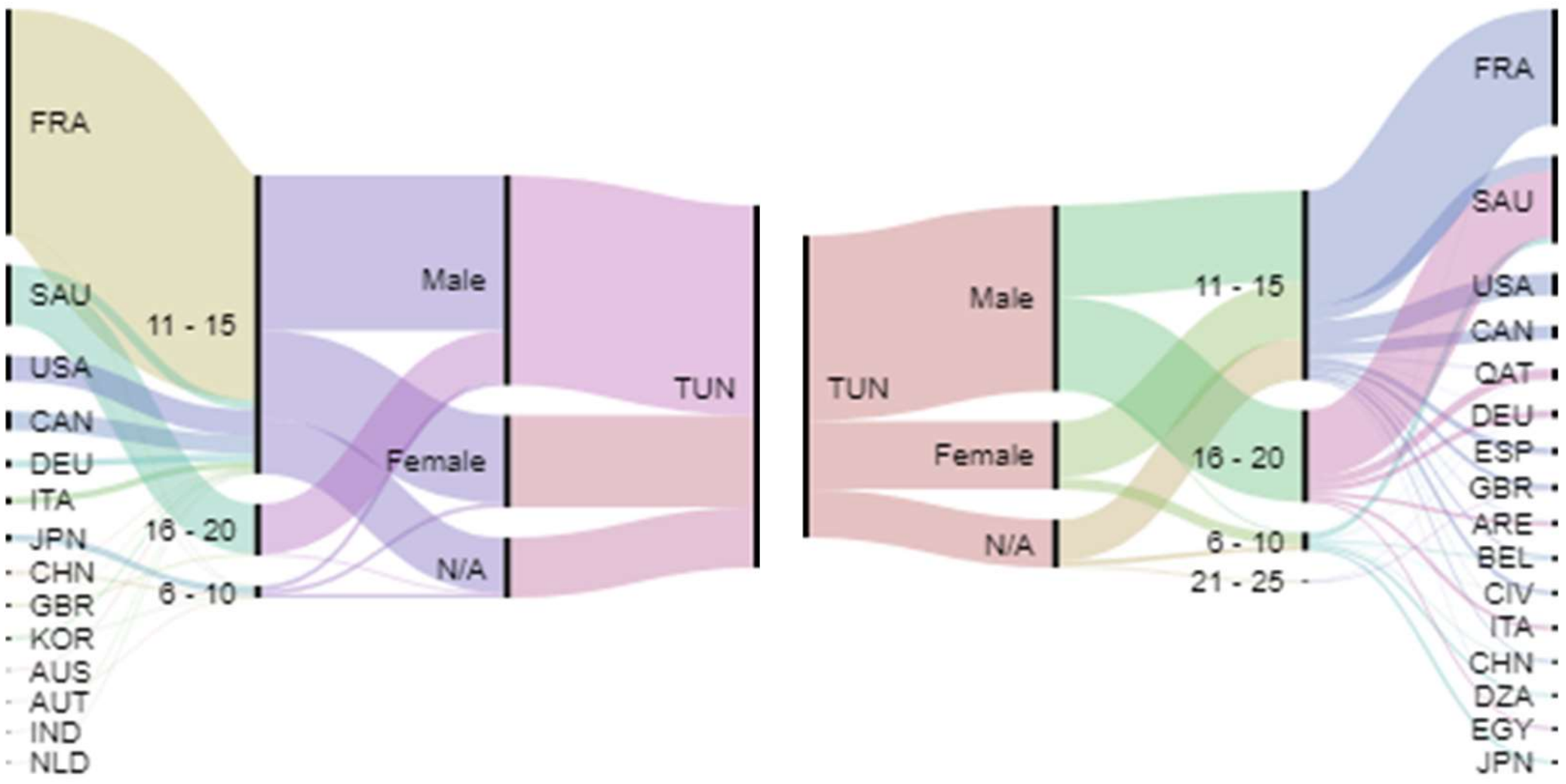

United Arab Emirates:

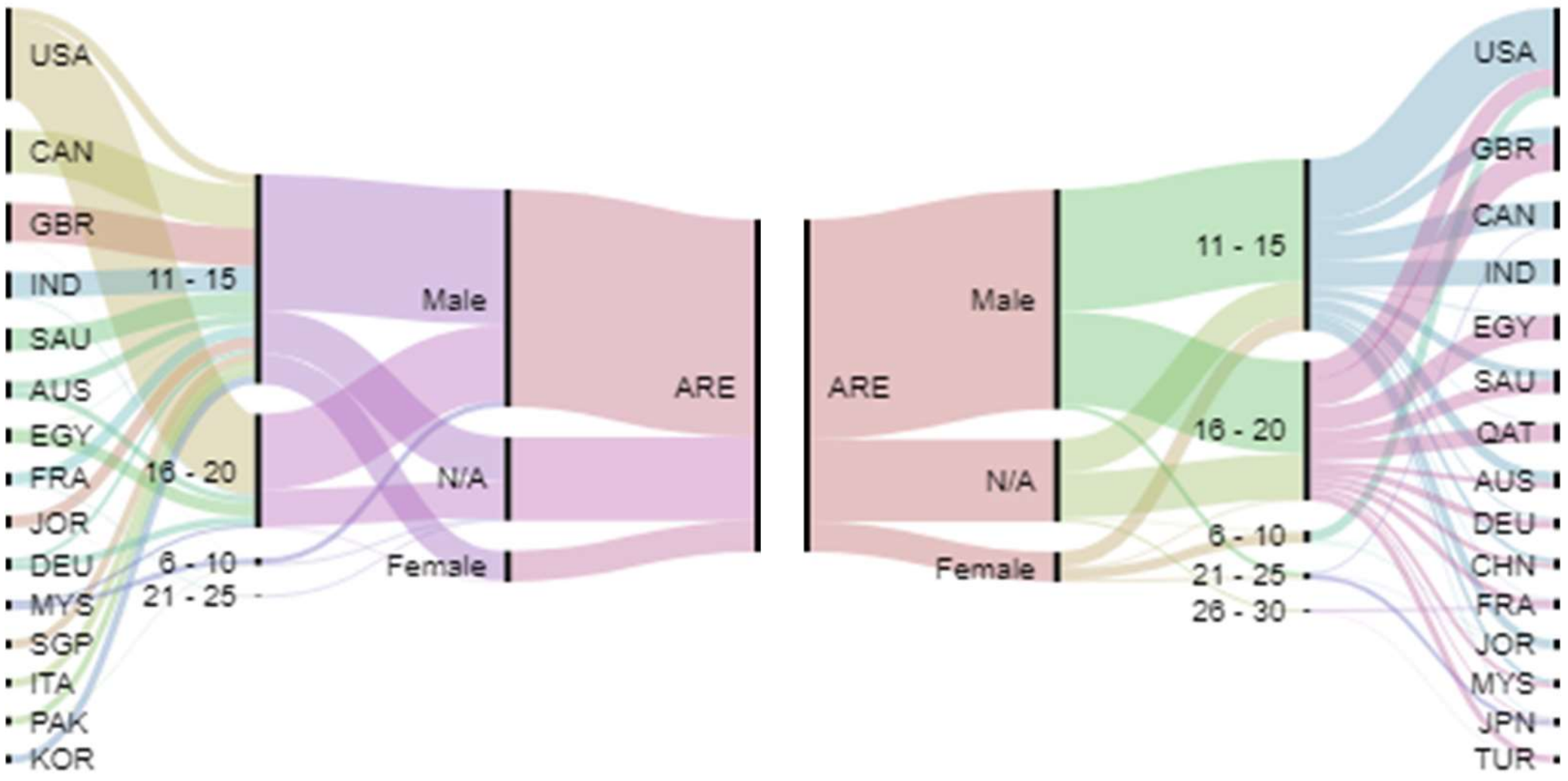

Algeria:

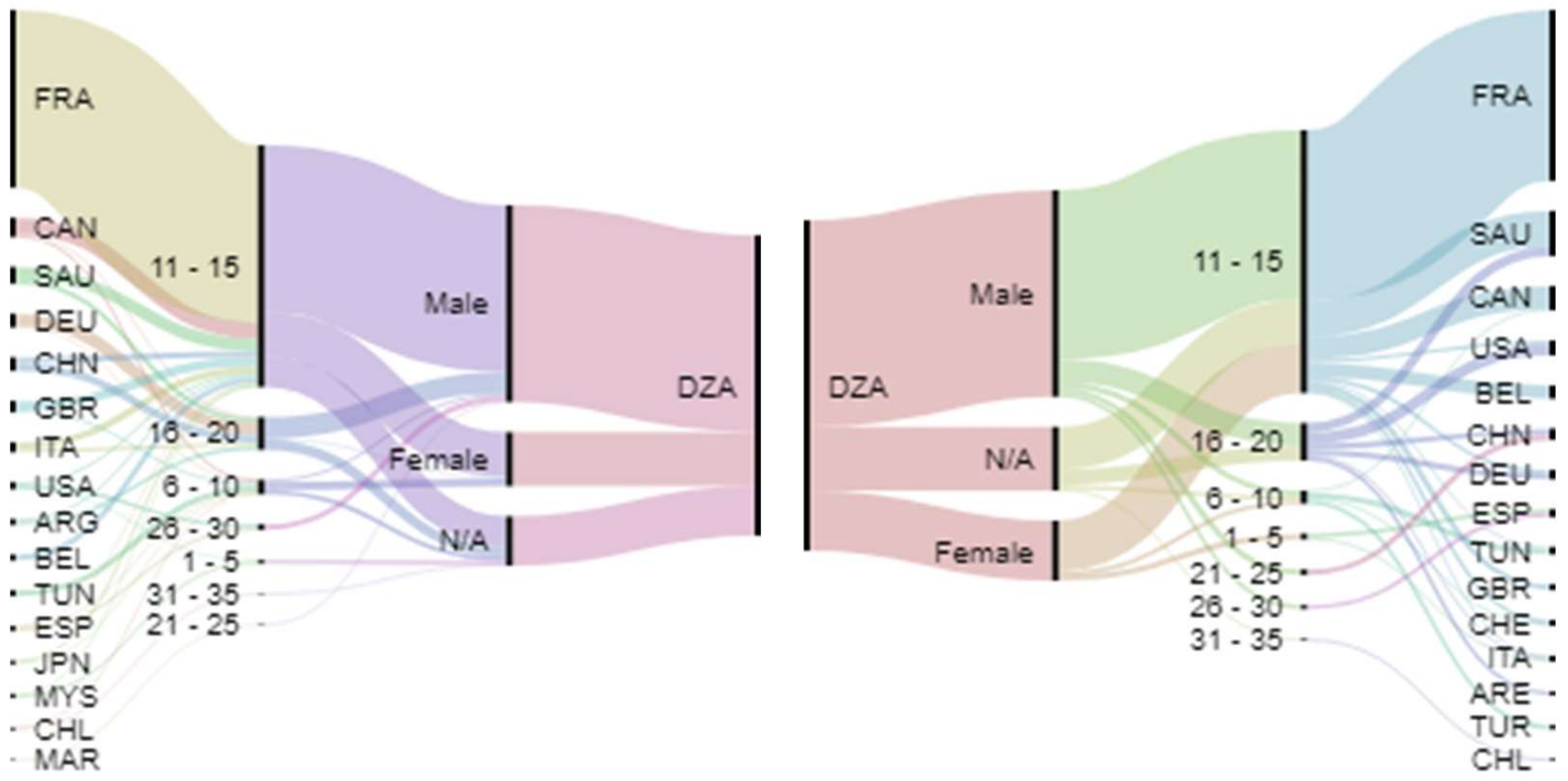

Lebanon:

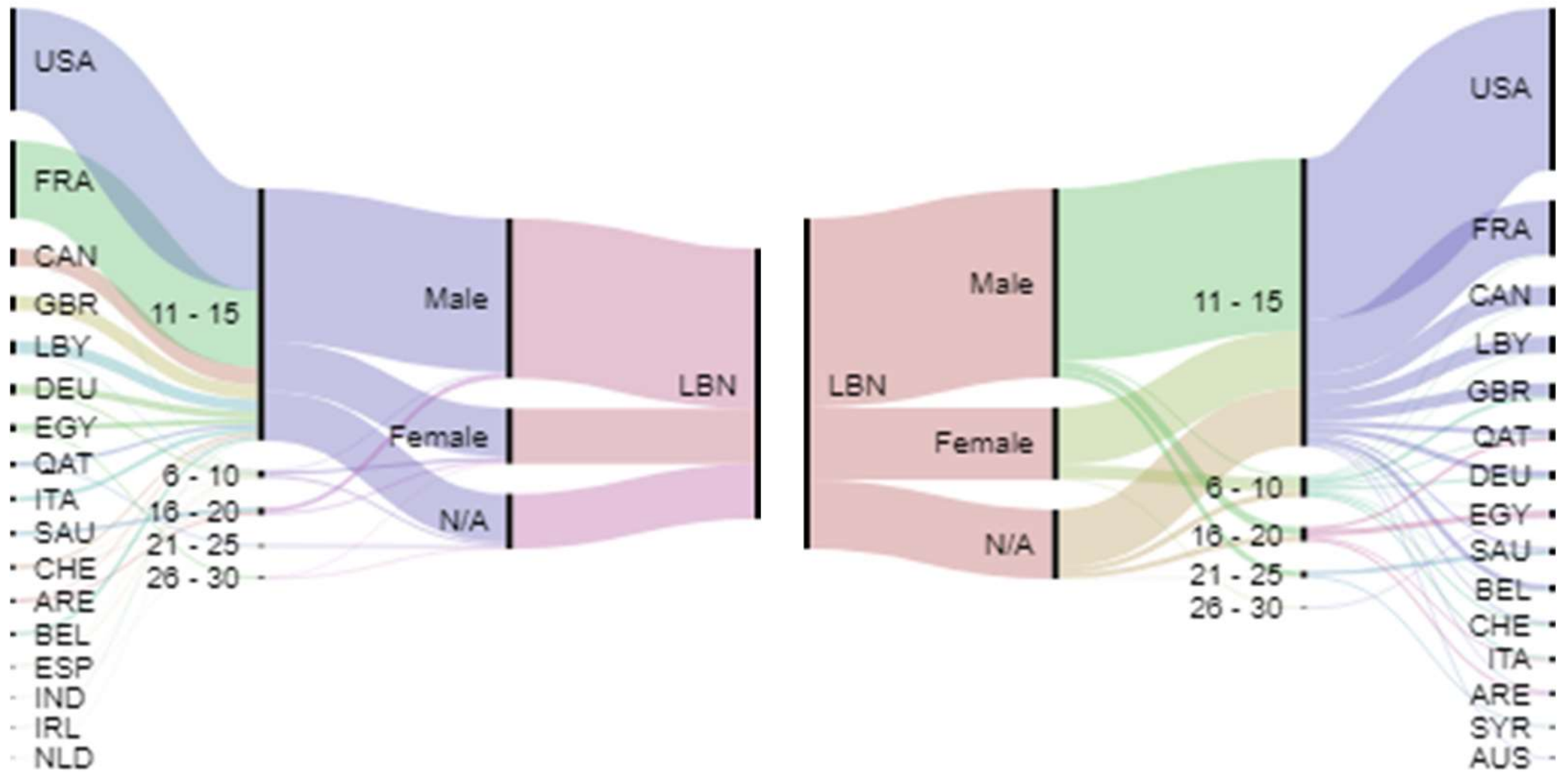

Morocco:

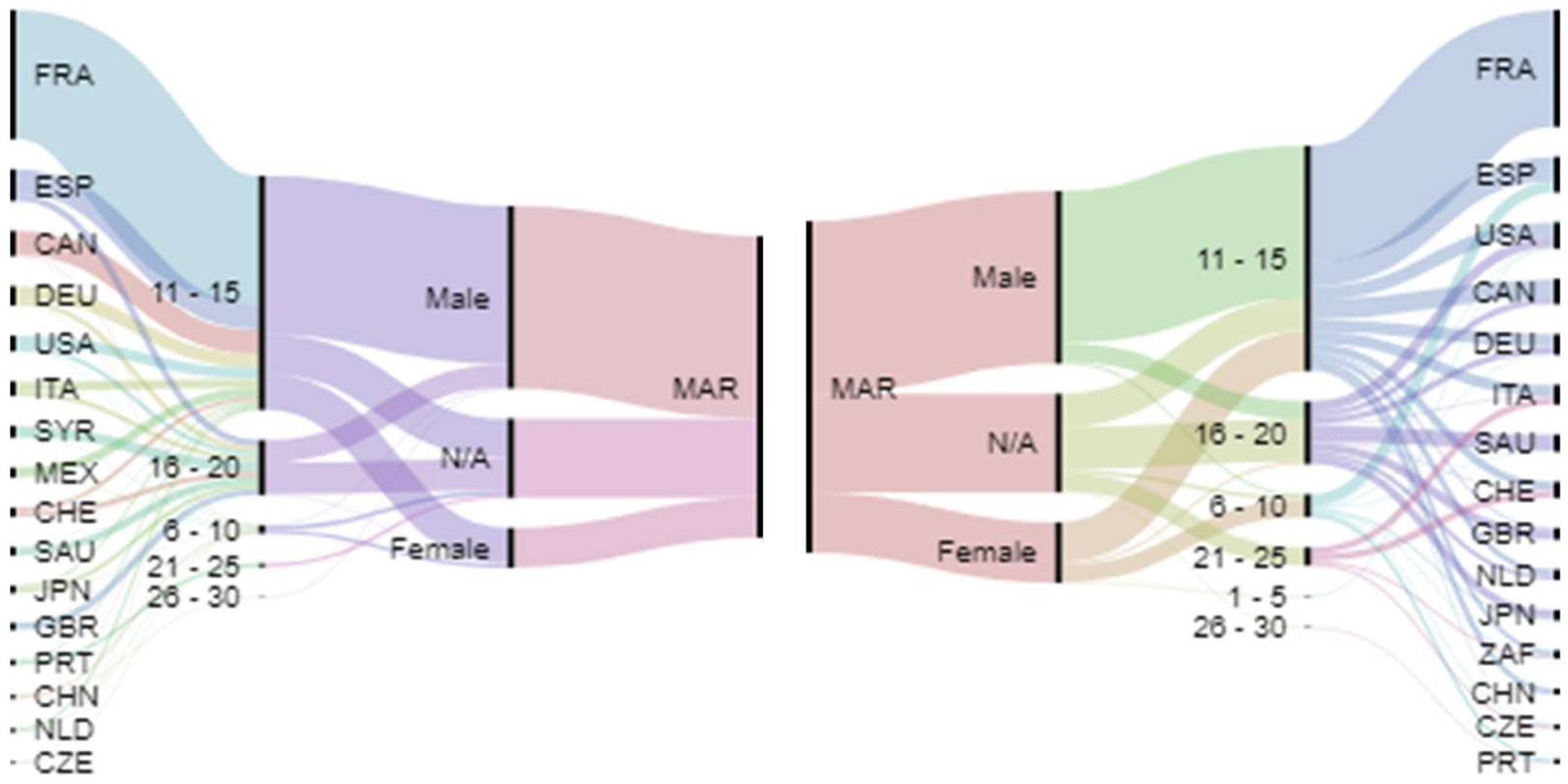

Qatar:

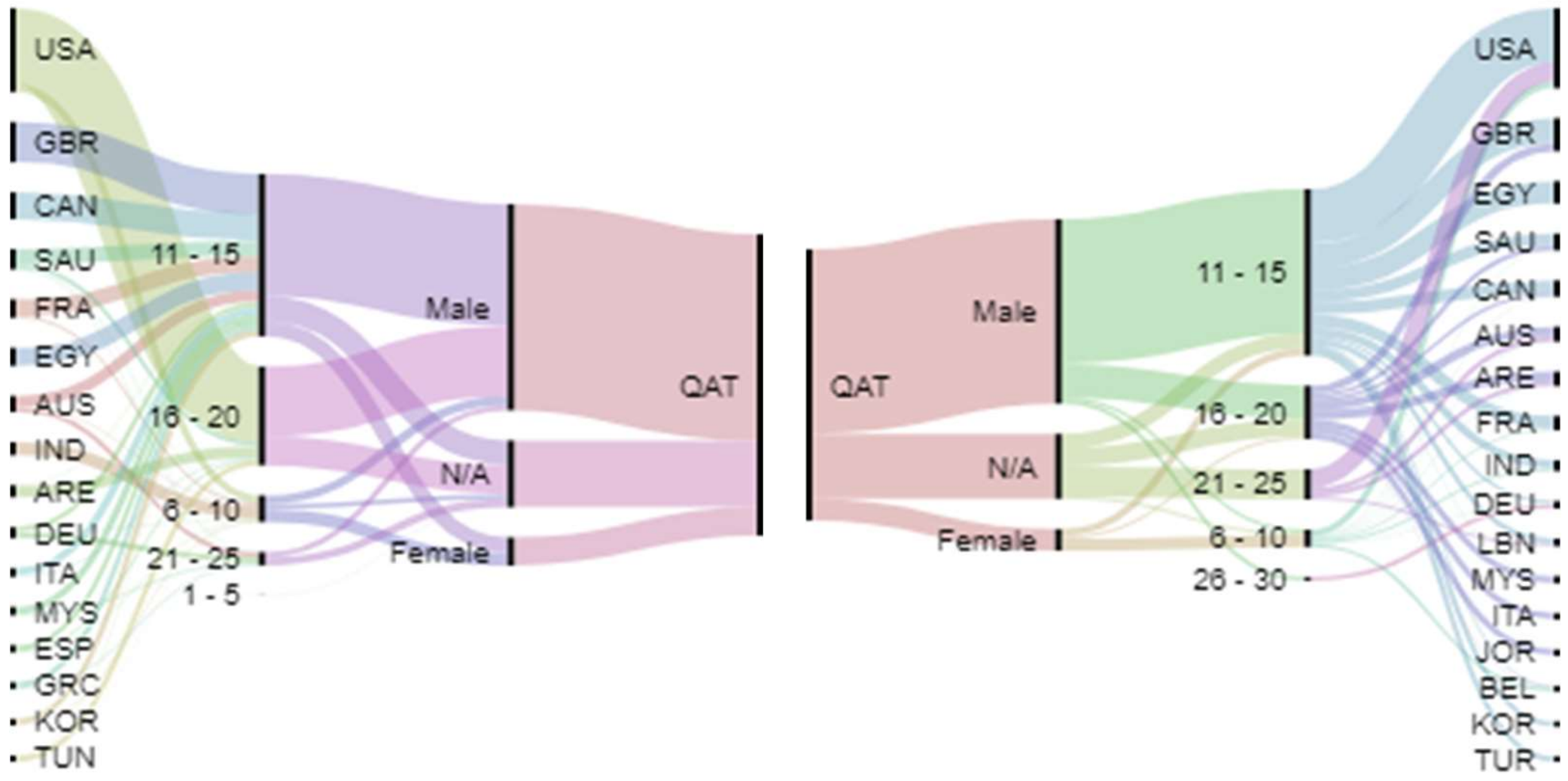

Iraq:

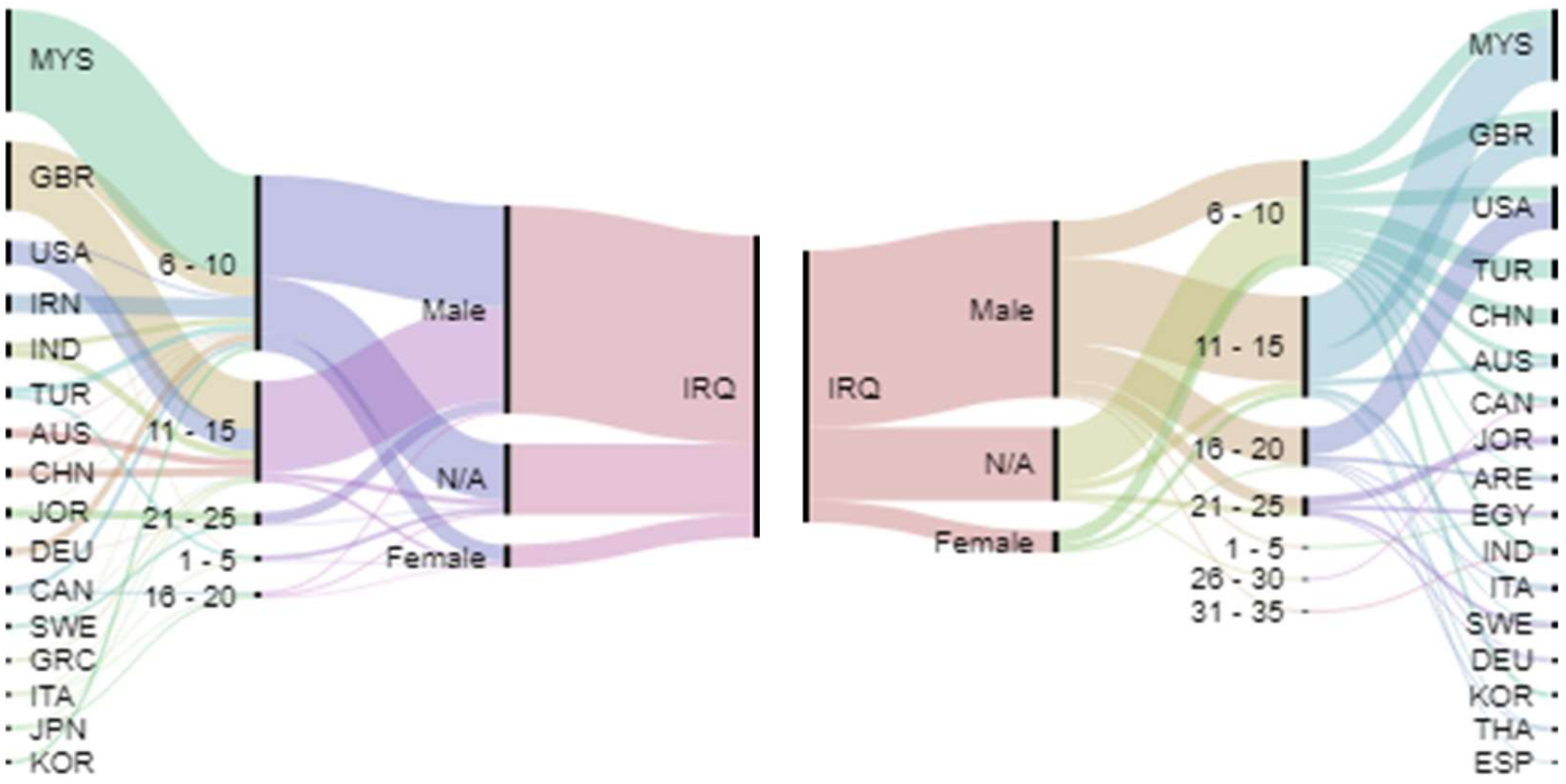

Jordan:

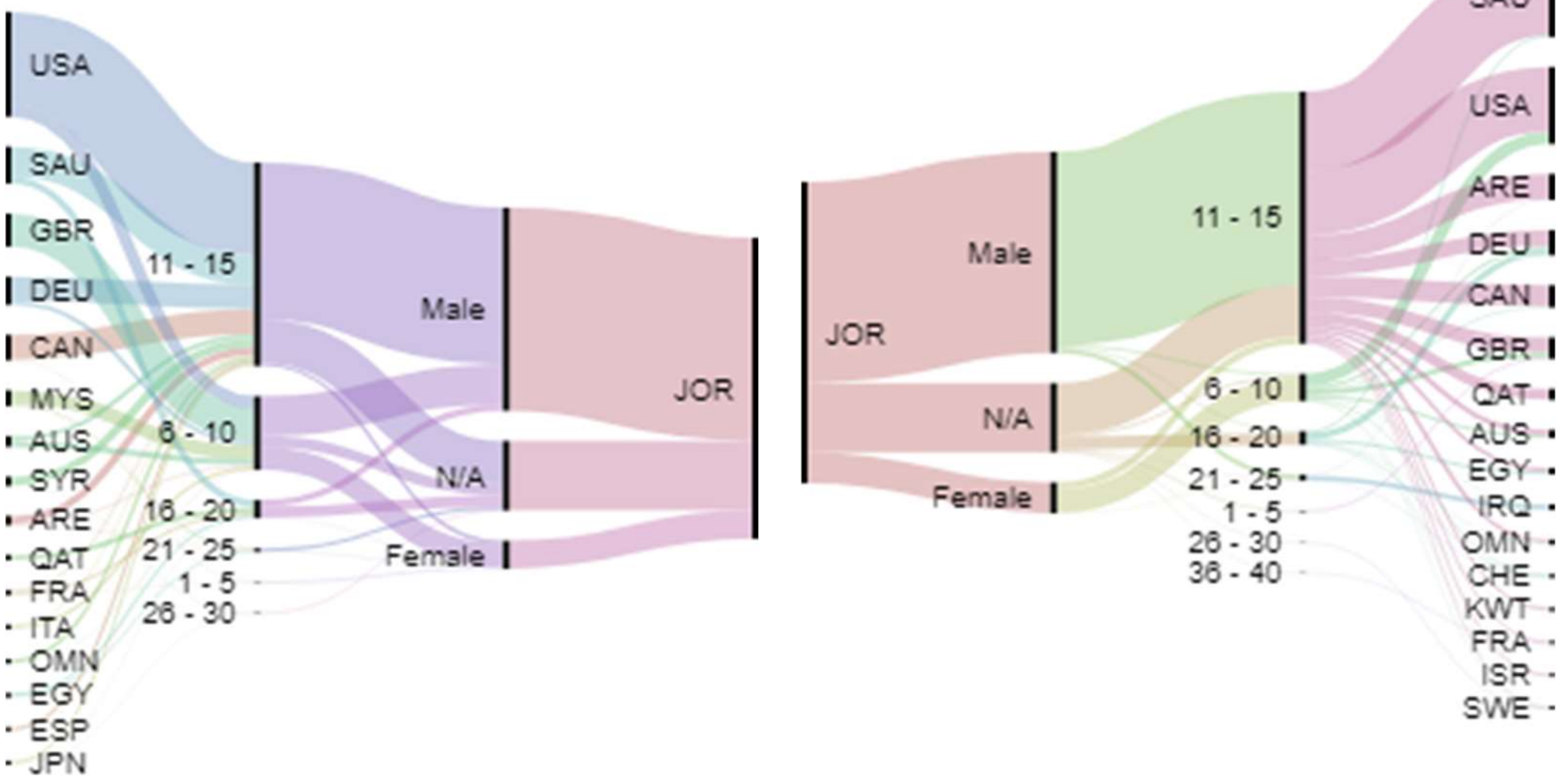